\documentstyle[12pt,epsfig,epstopdf,subfigure,color,amsfonts,amssymb]{article}
\newcommand{\beq}{\begin{equation}}
\newcommand{\eeq}{\end{equation}}
\newcommand{\bea}{\begin{eqnarray}}
\newcommand{\eea}{\end{eqnarray}}
\newcommand{\al}{\alpha}

\newcommand{\vep}{\varepsilon}
\newcommand{\ep}{\epsilon}

\newcommand{\der}{\partial}
\newcommand{\D}{\nabla}
\newcommand{\m}{\mu}
\newcommand{\n}{\nu}
\newcommand{\F}{{\cal{F}}}
\newcommand{\nn}{\nonumber}

\begin{document}

\baselineskip=15.5pt \pagestyle{plain} \setcounter{page}{1}
%
\begin{titlepage}

\vskip 0.8cm

\begin{center}

{\Large \bf  DIS off glueballs from string theory: the role of the
chiral anomaly and the Chern-Simons term}

\vskip 1.cm

{\large {{\bf Nicolas Kovensky}{\footnote{\tt
nico.koven@fisica.unlp.edu.ar}}, {\bf Gustavo Michalski}{\footnote{\tt
michalski@fisica.unlp.edu.ar}}, {\bf and Martin
Schvellinger}{\footnote{\tt martin@fisica.unlp.edu.ar}}}}

\vskip 1.cm

{\it Instituto de F\'{\i}sica La Plata-UNLP-CONICET \\ and \\
Departamento de F\'{\i}sica, Facultad de Ciencias Exactas,
Universidad Nacional de La Plata. \\
Calle 49 y 115, C.C. 67, (1900) La Plata, Buenos Aires, Argentina.} \\

\vspace{1.cm}

{\bf Abstract}

\vspace{1.cm}

\end{center}

We calculate the structure function $F_3(x,q^2)$ of the hadronic
tensor of deep inelastic scattering (DIS) of charged leptons from
glueballs of ${\cal {N}}=4$ SYM theory at strong coupling and at
small values of the Bjorken parameter in the gauge/string theory
duality framework. This is done in terms of type IIB superstring
theory scattering amplitudes. From the AdS$_5$ perspective, the
relevant part of the scattering amplitude comes from the
five-dimensional non-Abelian Chern-Simons terms in the $SU(4)$
gauged supergravity obtained from dimensional reduction on $S^5$.
From type IIB superstring theory we derive an effective Lagrangian
describing the four-point interaction in the local approximation.
The exponentially small regime of the Bjorken parameter is
investigated using Pomeron techniques.

\noindent

\end{titlepage}

\newpage

{\small \tableofcontents}

\newpage

\section{Introduction}

The holographic dual description of deep inelastic scattering (DIS)
of charged leptons from glueballs in the ${\cal {N}}=4$ $SU(N)$ SYM
theory with an IR cutoff has been proposed by Polchinski and
Strassler in \cite{Polchinski:2002jw}. In the planar limit, at
strong 't Hooft coupling ($1 \ll \lambda \ll N$), ${\cal {N}}=4$
$SU(N)$ SYM theory is dual to type IIB supergravity on AdS$_5 \times
S^5$. The compactification of type IIB supergravity on $S^5$ leads
to the maximally supersymmetric five-dimensional supergravity with
gauged $SU(4)$ symmetry
\cite{Kim:1985ez,Gunaydin:1984qu,Pernici:1985ju,
Gunaydin:1985cu,Freedman:1998tz}, and also there are Kaluza-Klein
modes. This dimensional reduction induces a five-dimensional
Chern-Simons term \cite{Gunaydin:1984qu,Pernici:1985ju,
Gunaydin:1985cu}, related to the chiral anomaly in the dual
${\cal{N}}=4$ SYM theory
\cite{Witten:1998qj,Freedman:1998tz,Bilal:1999ph}. In the
calculation of the hadronic tensor the chiral anomaly is reflected
in the fact that it appears a structure function $F_3$.

In the study of DIS in terms of the AdS/CFT correspondence there are
a few distinct parametric regions which depend on the relation
between the Bjorken parameter $x$ and the 't Hooft coupling of the
gauge theory, $\lambda=g_{YM}^2 \, N$, where $g_{YM}$ is the
coupling of the gauge theory. For the parametric region where
$1/\sqrt{\lambda} \ll x < 1$ the process is well described in terms
of type IIB supergravity. For $\exp{(-\sqrt{\lambda})} \ll x \ll
1/\sqrt{\lambda}$ excited strings are produced, therefore it is
necessary to consider type IIB superstring theory scattering
amplitudes in the holographic dual description. For exponentially
small values of $x$, diffusion effects become important and Pomeron
techniques can be used.

It is also possible to go beyond the tree-level approximation using
type IIB supergravity. Particularly in reference
\cite{Jorrin:2016rbx} $1/N^2$ corrections to DIS of charged leptons
off glueballs at strong coupling have been obtained, which
correspond to a DIS process where there are two-hadron final states.
Cutkosky rules allow us to calculate the imaginary part of an
amplitude by considering scattering amplitudes of the incoming and
outgoing states into all possible on-shell states. The result of
that calculation is very interesting, namely: the large $N$ limit
and the limit in which the momentum transfer of the virtual photon
is much larger than the IR cutoff do not commute. This indicates
that in the high energy limit two-particle intermediate states (in
terms of the Cutkosky rules) give the leading
contribution\footnote{In the paper \cite{Gao:2014nwa} also $1/N^2$
corrections have been considered. However, they have used an
effective model given by a scalar-vector Lagrangian, which has a
very small number of modes and interactions among them in comparison
with the actual possible field fluctuations of type IIB supergravity
which we have included in our paper \cite{Jorrin:2016rbx}.}.

Moreover, the holographic dual description of DIS from flavor
Dp-brane models has been carried out very successfully. Among the
interesting results, it is worth emphasizing that holographic dual
dynamical mesons show universal properties for the structure
functions \cite{Koile:2011aa,Koile:2013hba}. This is particularly
important because it should hold for scalar and polarized vector
meson structure functions for QCD itself, at least in the large $N$
limit\footnote{It could also hold for the first sub-leading term in
the $1/N$ expansion.}. The relevance of this comes from the fact
that the discovery of properties such as relations among the
structure functions (for example those similar to the Callan-Gross
relation) provides essential information about the internal
structure of hadrons, which can be helpful in order to study other
scattering processes. In addition, universal behavior suggests deep
underlying connections among different confining relativistic
quantum field theories. In this work we find new Callan-Gross type
relations for the antisymmetric structure function $F_3(x, q^2)$.

For scalar and polarized vector mesons new and very interesting
developments have been done in
\cite{Koile:2011aa,Koile:2013hba,Koile:2014vca}. Then, by using
these results for mesons a comparison with lattice QCD
data\footnote{Lattice QCD results of the first moments of the pion
and rho meson structure functions are presented in reference
\cite{Best:1997qp}.} has been carried out, finding good agreement
(within accuracy of $10\%$ or better) for an overall fitting of the
first three moments of the $F_2$ structure function of the pion, and
(within $21\%$ or better) for the first three moments of the $F_1$
structure function of the $\rho$-meson \cite{Koile:2015qsa}. These
calculations have been extended to one-loop level type IIB
supergravity for the D3D7-brane system, finding an impressive
improvement with respect to the tree-level results, now fitting
lattice QCD data within $1.27\%$ (or better) for the first three
moments of $F_2$ of the pion \cite{Kovensky:2016ryy}.

While most of the investigations outlined above concern the
calculation of the symmetric structure functions $F_1(x, q^2)$ and
$F_2(x, q^2)$, in the present work the interest is focused on the
antisymmetric structure function $F_3(x, q^2)$. We consider DIS of
charged leptons from glueballs in the ${\cal{N}}=4$ $SU(N)$ SYM
theory with an IR cutoff energy scale $\Lambda$, and describe it in
terms of its string theory dual model. It is interesting to recall
the origin of the antisymmetric structure functions which appear in
the hadronic tensor in this gauge theory. ${\cal{N}}=4$ $SU(N)$ SYM
theory has an $SU(4)_R$ R-symmetry group. The field content of the
gauge theory includes six real scalars transforming in the
representation {\bf 6}, and also there are four complex Weyl spinors
transforming in the fundamental representation of the R-symmetry
group with the chirality part $(0, 1/2)$ in the {\bf 4} and $(1/2,
0)$ in the {\bf 4$^*$} \cite{Ferrara:1998ej}. This $SU(4)_R$
symmetry is anomalous, {\it i.e.} it is broken at quantum level. The
anomaly can be calculated exactly at one-loop level, being the
corresponding Feynman diagram the one with three external points,
connected by three chiral fermion propagators. This is the so-called
triangle Feynman diagram, which is related to the three-point
function. The precise value of the chiral anomaly obtained
perturbatively from ${\cal{N}}=4$ $SU(N)$ SYM theory is
\cite{Freedman:1998tz,Schreier:1971um,Freedman:1992tz}
\bea
\frac{\partial}{\partial z_\rho}<J_\mu^A(x) J_\nu^B(y)
J_\rho^C(z)>_- =-\frac{N^2-1}{48 \pi^2} \, i \, d^{ABC} \,
\varepsilon_{\mu\nu\alpha\beta} \, \frac{\partial}{\partial
x_\alpha} \, \frac{\partial}{\partial x_\beta} \, \delta^{4}(x-z) \,
\delta^{4}(y-z) \, , \label{anomaly}
\eea
where the subindex minus indicates the abnormal piece of the
three-point function, {\it i.e.} the one which leads to the chiral
anomaly \cite{Freedman:1998tz}. $\varepsilon$ is the Levi-Civita
symbol, $d^{ABC}$ and $f^{ABC}$ are the $SU(4)_R$ symmetry group
symbols defined by Tr$(T^A T^B T^C) \equiv \frac{1}{4} (i
f^{ABC}+d^{ABC})$, where $T^A$ are hermitian generators of
$SU(4)_R$, which are normalized as Tr$(T^A T^B)=\frac{1}{2}
\delta^{AB}$. Considering the minimal coupling $\int d^4x J_\mu^A(x)
A^{A, \mu}(x)$, where $A^{A, \mu}(x)$ are background fields,
equation (\ref{anomaly}) can be rewritten as an operator equation
\bea
(D^\mu J_\mu(x))^A = \frac{N^2-1}{96 \pi^2} i \, d^{ABC} \,
\varepsilon^{\mu\nu\rho\sigma} \, \frac{\partial}{\partial x^\mu}
\left(A_\nu^B
\partial_\rho A_\sigma^C + \frac{1}{4} f^{CDE} A_\nu^B A_\rho^D
A_\sigma^E \right)
 \, . \label{anomaly-op-eq}
\eea
This anomaly is reflected in the bulk theory in a very nice way,
namely: since the global boundary $SU(4)_R$ symmetry corresponds to
a gauge $SU(4)$ symmetry in the AdS$_5$, the corresponding action in
the bulk is not gauge invariant \cite{Witten:1998qj}. It can be
easily seen by looking at the gauge sector of the action in AdS$_5$,
which after dimensional reduction of type IIB supergravity on the
five-sphere leads to the maximal $SU(4)$ gauged supergravity on
AdS$_5$. The action of this supergravity contains Chern-Simons term,
thus it is not gauge invariant. Moreover, the AdS/CFT correspondence
calculation shows the matching with the chiral anomaly of the
boundary theory \cite{Witten:1998qj,Freedman:1998tz,Bilal:1999ph}.
Let us recall how this works. The starting point is type IIB
supergravity in ten dimensions. In fact if one considers type IIB
superstring theory it turns out that the $1/N^2$ corrections only
come from the Kaluza-Klein modes arising from the dimensional
reduction on $S^5$, {\it i.e.} the $N^2-1$ overall factor in the
chiral anomaly\footnote{This was suggested by Witten
\cite{Witten:1998qj} and a more complete AdS/CFT calculation was
done by Bilal and Chu \cite{Bilal:1999ph}.}. After dimensional
reduction on $S^5$ it leads to the action for the $SU(4)$ gauge
fields $A_m^A(x, z)$
\bea
S_{5d}[A]=\int d^5x \, \left[\sqrt{-g_{AdS_5}} \, \frac{1}{4
g_{SG}^2} \, F^A_{m n} F^{A, m n} + \frac{i \, \kappa}{96 \pi^2}
(d^{ABC} \varepsilon^{mnopq} \, A^A_m \partial_n A_o^B
\partial_p A_q^C + \dots )\right] \, , \label{S5d0}
\eea
where $\kappa$ is an integer and we set $R=1$. Both the $SU(4)$
gauged supergravity coupling $g_{SG}$ as well as $\kappa$ are fixed
in terms of the boundary theory R-current correlators which are
exactly known \cite{Freedman:1998tz}. Parentheses in the action
(\ref{S5d0}) indicate the Chern-Simons term. Notice that Latin
indices stand for AdS$_5$ coordinates, while Greek indices denote
the boundary gauge theory coordinates. The Chern-Simons term are
proportional to the $SU(4)$ symmetric symbol $d^{ABC}$. Thus, this
is the origin of the quantum chiral anomaly in the dual ${\cal
{N}}=4$ $SU(N)$ SYM theory. From the Chern-Simons term above a
three-point interaction in AdS$_5$ is derived, which leads to the
three-point R-symmetry current correlator by using the AdS/CFT
correspondence, obtaining the following equation
\bea
(D^\mu J_\mu(x))^A = \frac{i \, \kappa}{96 \pi^2} \, d^{ABC} \,
\varepsilon^{\mu\nu\rho\sigma} \, \frac{\partial}{\partial x^\mu}
\left(A_\nu^B \partial_\rho A_\sigma^C + \frac{1}{4} f^{CDE} A_\nu^B
A_\rho^D A_\sigma^E \right)
 \, , \label{sugra-anomaly-op-eq}
\eea
where we have considered the boundary values of the bulk gauge
fields of the five-dimensional $SU(4)$ gauged supergravity,
$A_\mu^A(x) \equiv \lim_{z \rightarrow 0} A_\mu^A(x, z)$, which are
sources for the boundary theory global $SU(4)_R$ symmetry currents
$J_\mu^A(x)$. By matching equation (\ref{sugra-anomaly-op-eq}) to
equation (\ref{anomaly-op-eq}) it leads to $\kappa=N^2-1$. In
addition, the two-point R-symmetry current correlator fixes
$g_{SG}=4 \pi/N$. This indicates that in terms of the Witten's
diagrams the leading contributions from both terms in the action
(\ref{S5d0}) come with the same factor $N^2$.

Now, let us explain the consequences of the Chern-Simons term for
the calculation of the hadronic tensor of a scalar glueball in terms
of the gauge/string theory duality\footnote{There is a previous
calculation of the $F_3(x, q^2)$ structure function
\cite{Hatta:2009ra}, however this only contains a heuristic
five-dimensional approach and it has been done for spin-1/2
hadrons.}. The cubic part of the Chern-Simons term implies that in
the holographic calculation of the hadronic tensor, at small values
of the Bjorken parameter, the propagation of an $U(1) \subset
SU(4)_R$ gauge field in the $t$-channel is allowed. In the general
Lorentz covariant tensor decomposition of the current-current
correlator (which enters the definition of the hadronic tensor) this
term generates a tensor structure of the form
$\varepsilon^{\mu\nu\alpha\beta} \, q_\alpha \, P_\beta/(2 P \cdot
q)$, proportional to the $F_3(x, q^2)$ structure function. This
tensor is not invariant under parity transformations, thus a
non-conserving parity structure function appears in ${\cal {N}}=4$
$SU(N)$ SYM theory, and at small values of the Bjorken parameter we
find that this is of the same order as the $F_1(x, q^2)$ and $F_2(x,
q^2)$ structure functions. On the other hand, at larger values of
the Bjorken parameter we find that $F_3(x, q^2)$ is subleading in
comparison with $F_1(x, q^2)$ and $F_2(x, q^2)$.

Our findings are interesting since, to our knowledge, this is the
first result of the non-preserving parity structure function $F_3$
for a scalar hadron of ${\cal {N}}=4$ $SU(N)$ SYM theory. We have
obtained this in two different ways: firstly from a heuristic
calculation in five-dimensional $SU(4)$ gauged supergravity, and
then from a first principles type IIB superstring theory
calculation. Specifically, for small-$x$ values we obtain $F_3
\propto 1/x$, while for the exponentially small-$x$ region,
dominated by the $t$-channel Reggeized particle exchange, using
Pomeron techniques we find $F_3 \propto
(1/x)^{1-\frac{1}{2\sqrt{\lambda}}}$. Notice that for QCD in the
case of pure electromagnetic interaction $F_3=0$ since parity is
preserved (see for instance \cite{Anselmino:1994gn,Lampe:1998eu}).

~

This work is organized as follows. In the Introduction we describe
DIS in Yang-Mills theories and its description in terms of the
gauge/string duality. In Section 2 we show a heuristic derivation of
the effective Lagrangians from supergravity, which includes
symmetric contributions as well as antisymmetric contributions.
Then, in Section 3 we carry out a derivation of the effective action
directly from type IIB superstring theory which specifically leads
to the antisymmetric structure function $F_3$. This includes the
derivation of the Chern-Simons interaction from the superstring
theory scattering amplitude. In Section 4 we calculate the
antisymmetric structure function $F_3$ at small $x$ and comment on
the exponentially small-$x$ regime. In Section 5 we discuss our
calculations and results.

\subsection{Deep inelastic scattering in Yang-Mills theories}

Let us consider a charged lepton with four-momentum $k^\mu$
scattered from a hadron with four-momentum $P^{\mu}$ as
schematically shown in figure 1.a.
\begin{figure}[ht]
\centering
\begin{subfigure}[]{
\includegraphics[scale=0.35]{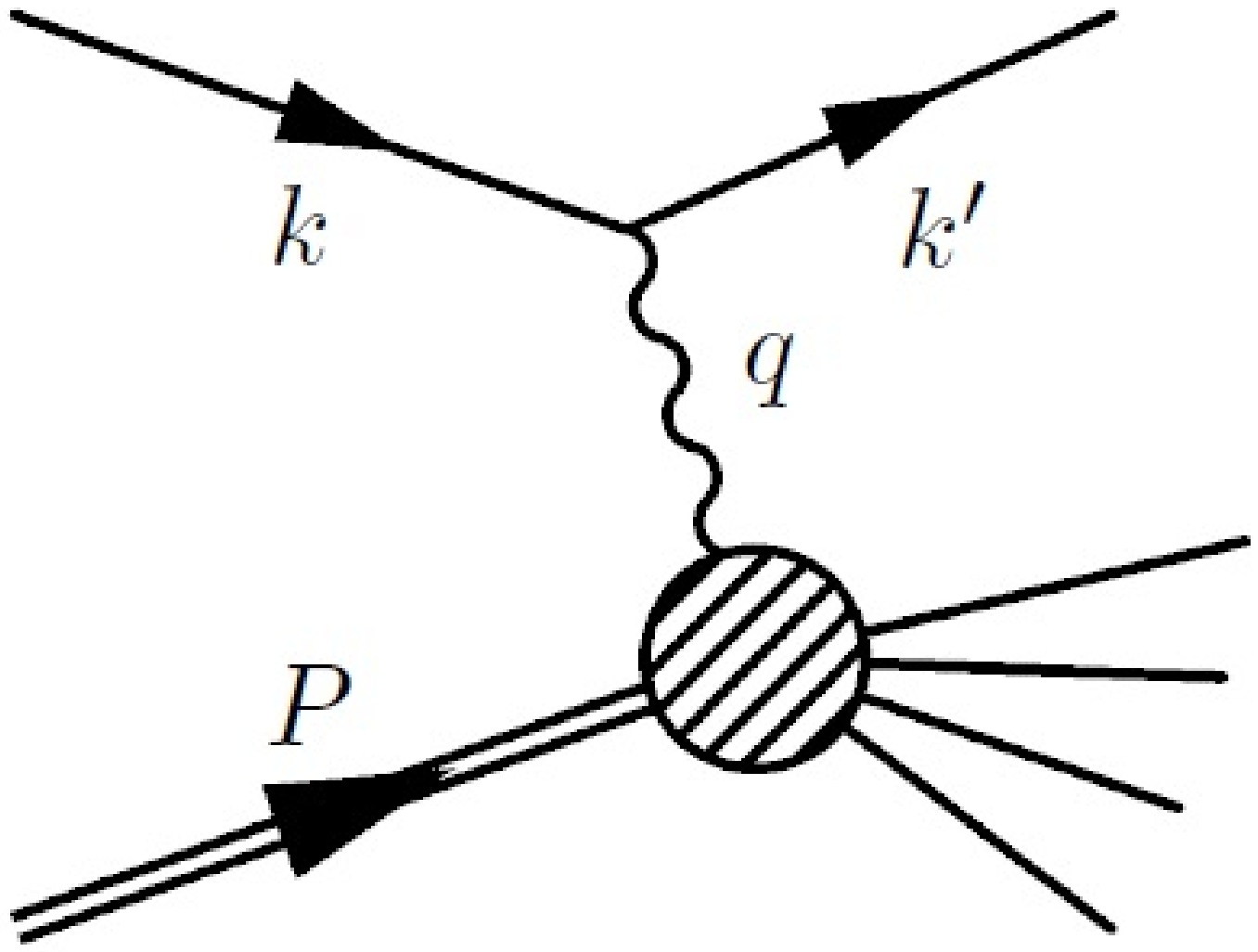}}
\end{subfigure}
\begin{subfigure}[]{
\includegraphics[scale=0.2]{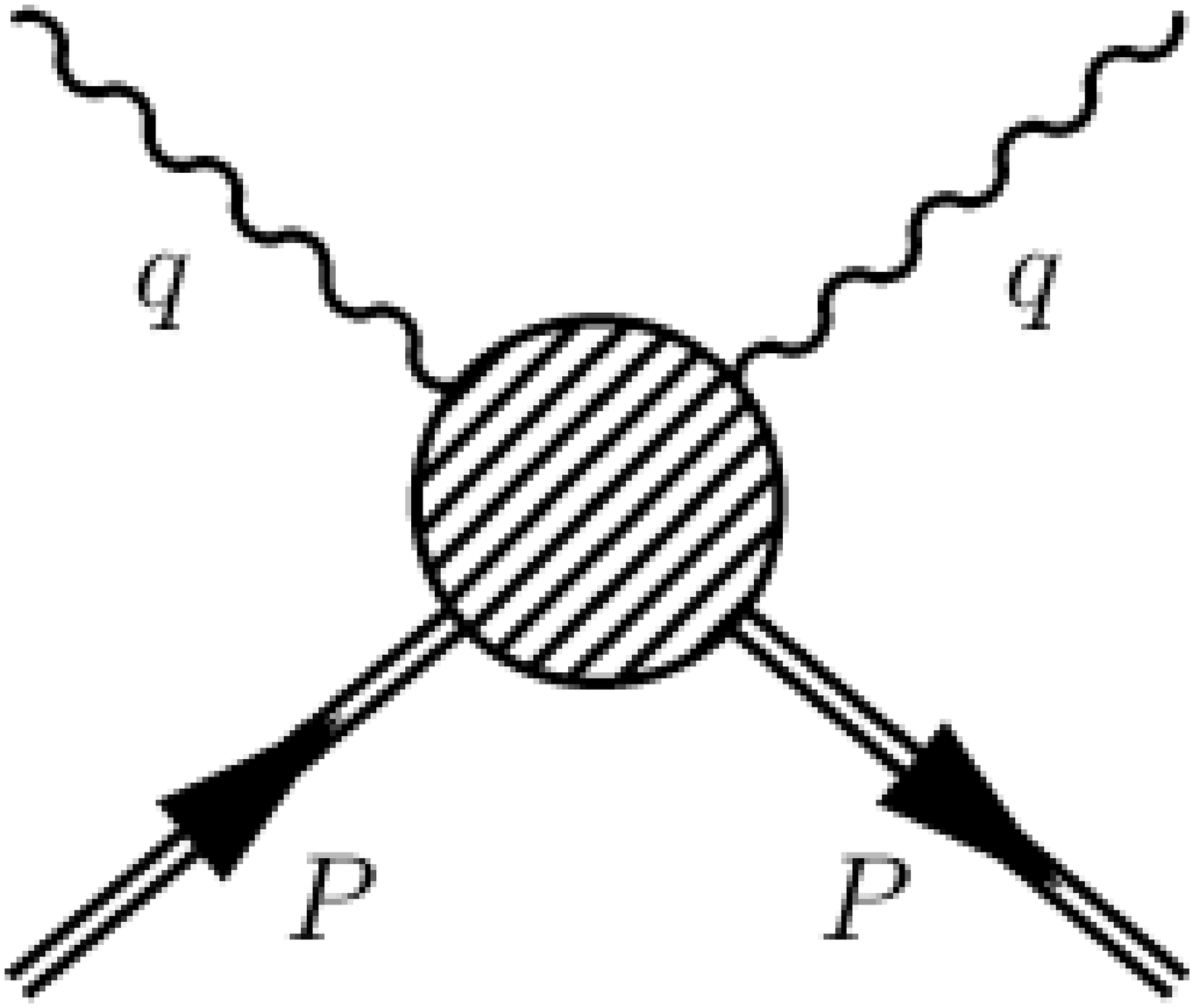}}
\end{subfigure}
\label{DISFCS} \caption{\small Schematic pictures of DIS (a) and
forward Compton scattering (b) processes.}
\end{figure}
The virtual photon carries four-momentum $q^\mu$. The associated
differential cross section is proportional to the $l_{\mu\nu}
W^{\mu\nu}$ contraction, where $l_{\mu\nu}$ is the leptonic tensor
calculated from perturbative QED. In contrast, the hadronic tensor
$W^{\mu\nu}$ involves soft processes, therefore it cannot be
calculated in perturbation theory. Its matrix elements are defined
as two-point functions of a commutator of electromagnetic currents
between the initial and final hadronic states with polarizations $h$
and $h'$
\begin{equation}
W^{\m\n}_{hh'}\equiv i \int d^{4}x\ e^{i q\cdot x} \langle P,h|
[J^\mu(x), J^\nu(0)] |P,h' \rangle \, .
\end{equation}
Time-reversal and translational invariance, hermicity restrictions
and Ward identities lead to several identities for the hadronic
tensor. As a result, it can be written as a sum of Lorentz covariant
tensor structures multiplied by the so-called structure functions,
which can be seen as functions of the virtual photon momentum
transfer $q$ and the Bjorken parameter
\begin{equation}
x=-\frac{q^2}{2P\cdot q} \, ,
\end{equation}
whose physical values belong to the range $0 \leq x \leq 1$. The DIS
regime corresponds to $q^2 \gg P^2$, keeping $x$ fixed. The hadronic
tensor can be decomposed in symmetric and antisymmetric terms under
$\m \leftrightarrow \n$. In particular, for scalar hadrons this
decomposition leads to
\begin{eqnarray}
W^{\m\n}(x,q^2) &=& W^{\m\n}_{\mathrm{S}}(x,q^2) + i \,
W^{\m\n}_{\mathrm{A}}(x,q^2) \label{Wdecomp}  \\
&=& \left(\eta^{\mu\nu} - \frac{q^\mu q^\nu}{q^2}\right) F_1(x,q^2)
- \left(P^\mu - \frac{P \cdot q}{q^2} q^\mu \right)
\left(P^\nu - \frac{P \cdot q}{q^2} q^\nu \right) \frac{F_2(x,q^2)}{P\cdot q} \nn \\
&& + i \, \vep^{\mu \nu \rho \sigma}
q_\rho P_\sigma \frac{F_3(x,q^2)}{2 P\cdot q} \nn \\
&=& \eta^{\mu\nu} F_1(x,q^2) + P^\mu P^\nu \frac{2x}{q^2} F_2(x,q^2)
- i \,\vep^{\m\n \rho \sigma} q_\rho P_\sigma \frac{x}{q^2}
F_3(x,q^2) + \dots \, . \nn
\end{eqnarray}
The last line of this equation has been rewritten in terms of $x$
and $q^2$. Also, dots indicate terms proportional to $q_\m$ which
can be omitted since after contraction with the leptonic tensor they
do not contribute to the DIS differential cross section. Notice,
that the third term would not be included if we had imposed parity
conservation. However, for ${\cal{N}}=4$ SYM theory a non-vanishing
$F_3$ structure function is expected even for {\it electromagnetic}
DIS.

Since there are contributions from soft processes to the DIS, the
structure functions cannot be obtained from pertubative SYM theory.
Fortunately, in certain parametric regimes DIS structure functions
can be obtained by using the gauge/string theory duality. DIS is
related to the forward Compton scattering (FCS) process through the
optical theorem. The related amplitude can be written in terms of a
tensor defined by the time-ordered expectation value of two
electromagnetic currents inside the hadron as follows
\begin{equation}
T^{\m\n}_{hh'}\equiv i \int d^{4}x e^{i q\cdot x} \langle P,h|
{\mathrm{\hat{T}}} \{J^\mu(x) J^\nu(0)\} |P,h' \rangle \, .
\end{equation}
The precise relation between the two tensors is given by the
following two equations
\begin{equation}
W^{\m\n}_{S} = 2 \pi \, {\mathrm{Im}}\left[T^{\m\n}_{S}\right] \ ,
\,\,\,\,\,\,\,\,\,\, W^{\m\n}_{A} = 2 \pi \,
{\mathrm{Im}}\left[T^{\m\n}_{A}\right] \, .
\label{relWT}
\end{equation}

The planar limit of ${\cal{N}}=4$ $SU(N)$ SYM theory is dual to a
particular solution of type IIB superstring theory, namely:
AdS$_5\times S^5$ background, with a constant dilaton and $N$ units
of the flux of the five-form field strength through $S^5$. It is
precisely in this context that the holographic dual picture of DIS
was developed in \cite{Polchinski:2002jw}. Moreover, the procedure
can be extended to other string theory dual models. In particular,
we will focus on the planar limit of ${\cal{N}}=4$ $SU(N)$ SYM
theory. In order to break conformal invariance and induce color
confinement the standard procedure requires to introduce an IR scale
$\Lambda$. Then, the hadron is represented by a state of mass $M
\sim \Lambda$. On the other hand, conformal symmetry is
asymptotically recovered in the UV limit, and at least at tree-level
the details of the IR structure are not important. An analogue to
the virtual photon of the DIS process is generated by gauging a
$U(1)$ subgroup of the $SU(4)_R$ R-symmetry group under which the
scalars and the fermions are charged. The conventional choice is to
use the $T^3=\textrm{diag}(1/2,-1/2,0,0)$ generator. This leads to
charges $\pm 1/2$ for two of the Weyl fermions and charge $1/2$ for
two complex scalars and the resulting gauge theory is anomaly free
since $d_{333}=0$ \cite{CaronHuot:2006te}. In this work we use the
three diagonal generators. It leads to a non-vanishing Chern-Simons
term in the dual supergravity description.

The explicit expression for the full non-Abelian conserved current
$J_\m^A$ (with $A = 1, \dots, 15$) in terms of the matter fields is
given in \cite{Hatta:2009ra,CaronHuot:2006te}. DIS of charged
leptons from glueballs in the large $N$ limit of ${\cal{N}}=4$
$SU(N)$ SYM theory has been described in detail in
\cite{Polchinski:2002jw}, in terms of the operator product expansion
(OPE) of the two electromagnetic currents inside the hadron. At weak
't Hooft coupling the OPE is dominated by single-trace twist-two
operators. However, at large coupling these operators develop large
anomalous dimensions and the main contribution to the OPE is given
by double-trace operators together with some specific protected
operators such as the energy-momentum tensor and the conserved
currents.

On the one hand, one can see that for moderate values of $x$ the
characteristics of the scattering are somewhat different in
comparison with QCD, namely: the relevant double-trace operators can
only create or annihilate an entire hadron, not being able to probe
its internal structure. This is related to the fact that particle
creation is suppressed in the bulk for $N \rightarrow \infty$.
One-loop level ($1/N^2$) corrections within this regime allow for
the photon to strike a secondary hadron from the surrounding cloud
of hadrons. On the other hand, for much smaller values of the
Bjorken parameter, in the $q^2 \rightarrow \infty$ limit the OPE is
dominated by the protected operators. This is in analogy with the
Pomeron description of the Regge regime of QCD. As we will see in
detail, this is dual to the $t$-channel graviton/gauge boson
exchange dominance in the bulk.

\subsection{Deep inelastic scattering and the  gauge/string duality}

The holographic dual model to the planar limit of ${\cal{N}}=4$ SYM
theory is given by a solution of type IIB supergravity on
AdS$_5\times S^5$, with radius $R$ and the metric \footnote{We use
the following conventions: $M, N,\dots=0,\dots, 9$ are the
ten-dimensional indices, $m, n,\dots=0,\dots, 4$ are AdS$_5$
indices, $\m,\n,\dots=0,\dots, 3$ are flat four-dimensional indices
and $a,b,\dots=1,\dots,5$ are $S^5$ indices.}
\begin{equation}
ds^2 =
\frac{R^2}{z^2}\left(\eta_{\m\n}dx^\m dx^\n+dz^2\right)+ R^2
d\Omega_5^2 \, .
\end{equation}
In terms of these coordinate the UV boundary is located at
$z\rightarrow 0$. The relation between the number of color degrees
of freedom, $N$, the 't Hooft coupling $\lambda$ of the gauge
theory, and the parameters of the string theory is given by
\begin{equation}
\frac{R^2}{\al'} = \sqrt{4 \pi \lambda} \ , \,\,\,\,\,\,\,\,\, g_s
\equiv g_{YM}^2 \, ,
\end{equation}
where $\alpha'=l_s^2$ is the string length and $g_s$ is the string
coupling.

The introduction of an IR scale $\Lambda$ in the gauge theory
corresponds to a cutoff in the small $z$ region. Since the details
of the IR are not important, we use an over-simplified deformation
known as the hard-wall model, in which the anti-de Sitter
description is assumed to be exactly valid up to the point
$z_0=1/\Lambda$. Since hadronic states at the boundary are dual to
normalizable modes in the bulk, by imposing Dirichlet boundary
conditions at this point leads to a restriction for the dual hadron
mass. In this work, we will focus on glueballs created by operators
which are dual to normalizable modes in the Kaluza-Klein (KK) tower
associated to the ten-dimensional dilaton field $\phi$. For example,
for the incoming mode the solution corresponding to a state created
by an operator with scaling dimension $\Delta$ has a KK mass
$R^{-2}\Delta(\Delta -4)$ from the point of view of the
five-dimensional theory. Thus, the ten-dimensional field is given by
\begin{equation}
\phi_i (x^\m, z, \Omega) = c_i \frac{\sqrt{P \Lambda}}{R^4} e^{i
P\cdot x} z^2 J_{\Delta-2}(P z) Y_{\Delta}(\Omega)\approx
\frac{c_i}{\Lambda R^4}e^{i P\cdot x}
\left(\frac{z}{z_0}\right)^{\Delta} Y_{\Delta}(\Omega),
\label{solphi}
\end{equation}
where in the last expression we have expanded near the boundary.
$c_i$ is some numerical normalization constant and
$Y_{\Delta}(\Omega)$ is a scalar spherical harmonic on the
five-sphere\footnote{The normalization condition is given in the
appendix of \cite{Polchinski:2001tt}. In their conventions the
spherical harmonic $Y_{\Delta}$ is normalized over the unit
five-sphere.}. On the other hand, the holographic dual of the
virtual photon is given by a non-normalizable mode of a gauge field
$A_m$ in the bulk\footnote{In this work we use the convention $A_m^3
\equiv A_m$.}. For the ingoing field, the solution to the associated
Einstein-Maxwell equations on AdS$_5$ and the corresponding boundary
conditions are
\begin{eqnarray}
&& A_\mu (x^\nu,z) = n_\mu e^{i q \cdot x} \, q z K_1 (qz) \, ,
\,\,\,\,\, A_z (x^\nu,z) = -i (n \cdot q) \, e^{i q \cdot x}
\frac{z^3}{R^2} K_0(qz) \, , \nonumber \\
&& A_\mu (x^\nu, z\rightarrow 0) = n_\mu \, e^{i q \cdot x} \ ,
\,\,\,\,\,\,\, A_z(x^\nu, z\rightarrow 0) = 0 \, . \label{solA}
\end{eqnarray}
We can set the transversal polarization condition $n \cdot q=0$. The
Bessel function of the second kind $K_1(qz)$ vanishes exponentially
as $q z$ increases in the bulk, which indicates that the interaction
must occur at $z_{int} \sim 1/q$, leading to a suppression factor
$(\Lambda^2/q^2)^{\Delta-1}$, at least when $x$ is not exponentially
small.

The gravity counterparts for the different parametric regimes
described above from the field theory viewpoint can be obtained by
looking at the center-of-mass energy. There is the following
parametric relation \cite{Polchinski:2002jw}
\begin{equation}
\tilde{s} \lesssim \frac{z_{int}^2}{R^2} \, s = \frac{1}{\sqrt{4 \pi
\lambda} \,  \al'} \left(\frac{1}{x}-1\right) \, ,
\end{equation}
where $s$ is the four-dimensional Mandelstam variable and
$\tilde{s}$ is its ten-dimensional counterpart. Thus, the IIB
supergravity description of the bulk dynamics corresponds to the
range $1 > x \gg \lambda^{-1/2}$ on the gauge field theory side. In
this case the leading amplitude of the dual FCS process is given by
an $s$-channel diagram in type IIB supergravity. In this parametric
regime the photon strikes the entire hadron. Then, the DIS structure
functions are obtained from the imaginary part of the two-point
current correlator by considering the on-shell propagator. In
contrast, when $x \ll \lambda^{-1/2}$ we see that $\al' \tilde{s}$
is order one, therefore the type IIB superstring theory dynamics
becomes relevant, and consequently the exchange of a Reggeized
graviton mode dominates. In the $\exp(-\lambda^{1/2}) \ll x\ll
\lambda^{-1/2}$ regime the interaction can be thought of as local,
thus it can be described in terms of an effective action deduced
from flat-space string theory scattering amplitudes. For the
smallest parametric region, {\it i.e.} when $x \leq
\exp(-\lambda^{1/2})$ diffusion effects in the radial direction
become important and the interaction cannot be considered local.
This region can be described in terms of the Pomeron
\cite{Polchinski:2002jw,Brower:2006ea}.

In the holographic picture, the ${\cal {N}}=4$ SYM R-symmetry group
corresponds to the isometry group of the five-sphere, $SU(4)\sim
SO(6)$. It can be gauged in order to construct a five-dimensional
gauged supergravity on AdS$_5$ \cite{Gunaydin:1985cu,Kim:1985ez}.
From the ten-dimensional perspective, the corresponding gauge fields
arise as perturbations of some particular fields which are expanded
in modes on $S^5$. The details of the five-dimensional reduction are
given in reference \cite{Kim:1985ez}. The excitations of the
graviton $h_{m a}$ and the Ramond-Ramond (RR) 4-form $a_{m a b c}$
with one index in the AdS$_5$ can be written as
\begin{equation}
h_{ma} = \sum_k B_m^{(k)}(x^n) \, Y_a^{(k)}(\Omega) \ ,
\,\,\,\,\,\,\,\,
a_{mabc} = \sum_k \tilde{B}_m^{(k)}(x^n) \,
\ep_{abc}^{\,\,\,\,\,\,\,\,de} \, \nabla_d Y_{e}^{(k)}(\Omega) \, ,
\end{equation}
where $\ep_{abcde}$ is the Levi-Civita tensor density on $S^5$,
$Y_a^{(k)}(\Omega)$ are vector spherical harmonics, where $k \geq 1$
label their corresponding $SU(4) \approx SO(6)$ representations,
while $B_m(x^n)$ and $\tilde{B}_m(x^m)$ are vector fields in
AdS$_5$. At the lowest level, $k=1$, the spherical harmonics
correspond to the $S^5$ Killing vectors $K_a^A$. After
diagonalization of the equations of motion associated with these
modes, the fifteen AdS$_5$ massless gauge fields arise as the
following linear combination
\begin{equation}
A_m^A \equiv B_m^A - \frac{16}{R}\tilde{B}_m^A \, .
\label{A-modes}
\end{equation}
The second contribution can be ignored in the supergravity
calculations and also in the construction of the effective action
that leads to the symmetric structure functions in the small-$x$
regime \cite{Polchinski:2002jw}. However, for the holographic dual
description of the antisymmetric structure functions the second
contribution of equation (\ref{A-modes}) must be included.

\section{Heuristic effective Lagrangian from supergravity}

In the high center-of-mass energy regime, {\it i.e. $x\ll1$}, the
holographic dual description of DIS in the bulk is given by the
exchange of excited strings states. In this situation it is
necessary to go beyond the supergravity description. Thus, it
requires considering string theory scattering amplitudes, which can
be expressed as the product of a pre-factor ${\cal{G}}(\al'; s, t,
u)$, which contains the $\al'$ dependence, and a kinematic factor
${\cal{K}}$. This amplitude is calculated in order to build an
effective Lagrangian from which the hadronic tensor can be
calculated after evaluating on the solutions of the fields in
AdS$_5$.

The effective Lagrangian may be obtained in a heuristic way by
analyzing the five-dimensional gauged supergravity diagram of the
photon-dilaton to photon-dilaton scattering at tree level, where the
leading diagram in the high energy limit is given by the
$t$-channel. This heuristic method was discussed in
\cite{Gao:2009se}, where non-forward Compton scattering amplitudes
for dilatons have been calculated.

The supergravity action on AdS$_5$, with indices $m, n = 0, ..., 4$,
can be written as
\bea
S_{5d} = \frac{1}{2\kappa^2_5}\int d^5x
\sqrt{-g_{AdS_5}}\left(-{\cal{R}}+\frac{1}{2}(\der_m\phi)^2
+\frac{1}{4}\left(F_{mn}^A\right)^2+\cdots \right) + S_{CS} \, .
\label{s5d}
\eea
In this section and in the following one we set $R=1$, thus
$2\kappa_5^2 = 16 \pi^2/N^2$. Also, $F^A_{mn}$ is the non-Abelian
field strength associated with the gauge fields, $\phi$ is the
dilaton and ${\cal{R}}$ the Ricci scalar which includes the graviton
$h_{mn}$. Dots include the kinematic terms of the fields not
relevant for our analysis, and also the interaction terms of the
type $(\phi\phi h)$, $(AAh)$ and $(A\phi\phi)$. The last factor
$S_{CS}$ is the Chern-Simons term defined in (\ref{S5d0}).

In this section we calculate the heuristic Lagrangian obtained from
the $\phi A \rightarrow \phi A$ scattering mediated by a graviton,
and show that it coincides with the one calculated in
\cite{Polchinski:2002jw} from closed string amplitudes. Then, we
will use the same techniques to calculate the effective Lagrangian
from which the leading contribution to the antisymmetric structure
function $F_3(x,q^2)$ can be obtained. The Lagrangian arises from
the $\phi A \rightarrow \phi A$ scattering with the exchange of a
gauge field in five-dimensional gauged supergravity on AdS$_5$.

\subsection{Symmetric contributions}

\begin{figure}[ht]
\begin{center}
\includegraphics[scale=0.8]{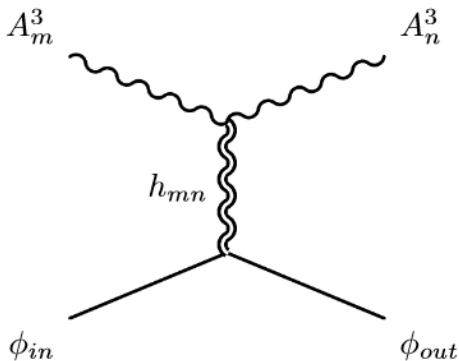}
\caption{\small Feynman diagram corresponding to the graviton
exchange contribution to the DIS (FCS) process for small values of
the Bjorken parameter $x$.} \label{Tgrav}
\end{center}
\end{figure}

The idea is to calculate the four-point scattering amplitude where
the ingoing and outgoing states are given by a dilaton $\phi$ dual
to the hadron, and a gauge field $A^3_m$ dual to the $U(1)$ current,
interacting throughout the exchange of an AdS$_5$ graviton. The
process is schematically shown in figure \ref{Tgrav}.
Notice that in reference \cite{Gao:2009se} only the AdS$_5$
components of the field decomposition have been considered, thus
ignoring the Kaluza-Klein modes coming from the $S^5$ because they
only contribute with a global constant. Given that the graviton
couples to the energy-momentum tensor $T_{mn}$, the scattering
amplitude in terms of the perturbations\footnote{We parameterize the
perturbations as $\Phi \to \Phi_0 + \sqrt{2}\kappa_5 \Phi$, thus neither
the energy-momentum tensor nor the propagator have $\kappa_5$
factors.} is given by
\begin{equation}
{\cal{A}}= \kappa_5^2 \int d^5x \, d^5x' \, T_{mn}^\phi(x) \,
G^{mnkl}(x,x') \, T_{kl}^A(x') \, , \label{AsymK5}
\end{equation}
where the AdS$_5$ graviton propagator in the high energy limit can
be expressed as \cite{D'Hoker:1999jc,Bartels:2009sc}
\begin{equation}
G^{mnlk}(x,x')=\left(g^{mk}g^{nl}+g^{ml}g^{nk} - \frac{2}{3} \,
g^{mn}g^{kl}\right) G_{\mathrm{grav}}(x,x') \, ,
\end{equation}
with $G_{\mathrm{grav}}(x,x')$ being some function that is not
relevant in the present case, while the dilaton and gauge field
energy-momentum tensors are given by
\bea
T_{mn}^\phi = (g_{m p}g_{n q}+g_{m q}g_{n p}-g_{mn}g_{pq})\der^p
\phi^*\der^q \phi \;, \qquad
T^A_{kl}=g^{pq}F_{kp}F_{lq}-\frac{1}{4}g_{kl}F_{pq}F^{pq} \, ,
\eea
respectively. The contraction of these three tensors leads to
\bea
&& T_{mn}^\phi(x) \, G^{mnlk}(x,x') \, T^A_{kl}(x') =
G_{{\mathrm{grav}}}(x,x') \, \times \nonumber \\
&& \,\,\,\,\,\,\,\,\,\,\,\,\,\,\,\,
\left[2\,(\der^k\phi^*(x)\der^l\phi(x)+\der^l\phi^*(x)\der^k\phi(x))
F_{kp}(x')F_{lq}(x')g^{pq}
+ \cdots \,\right] , \label{LheuS0}
\eea
where we only write the leading terms in the $s \to \infty$ and $t
\to 0$ limits.  After integration, this expression matches the index
structure of ${\cal{K}}|_{t\simeq0}$ of equation (2.38) of
\cite{Koile:2014vca}\footnote{In equation (2.38) of
\cite{Koile:2014vca} we have corrected several mistakes in equation
(82) of \cite{Polchinski:2002jw}.}.

Since the derivation of this section is heuristic, in order to
obtain the same action as in reference \cite{Polchinski:2002jw} we
must multiply by the factor
${\cal{G}}(\al',\tilde{s},\tilde{t},\tilde{u}) \, {\tilde {s}}^2$
included in the four-point string theory scattering amplitude, where
\begin{equation}
{\cal{G}}(\al',\tilde{s},\tilde{t},\tilde{u}) = - \frac{\al'^3}{64}
\frac{\Gamma\left(-\al'\tilde{s}/4\right)
\Gamma\left(-\al'\tilde{t}/4\right)
\Gamma\left(-\al'\tilde{u}/4\right)}{
\Gamma\left(1+\al'\tilde{s}/4\right)
\Gamma\left(1+\al'\tilde{t}/4\right)
\Gamma\left(1+\al'\tilde{u}/4\right)} \, .
\end{equation}
While at this level of derivation this is an {\it ad hoc} factor, it
naturally appears when considering the four-point string theory
scattering amplitude. It leads to the possibility of exchanging a
whole tower of excited string states. This factor is particularly
relevant because it leads to a finite contribution from equation
(\ref{AsymK5}) to the imaginary part of the scattering amplitude.
Thus, the effective action turns out to be
\beq
S_{{\mathrm{eff}}}^{{\mathrm{Sym}}} = \,
{\mathrm{Im}}\left[{\cal{G}}\tilde{s}^2\right] \ \kappa_5^2\ \int
d^5\Omega \sqrt{g_{S^5}} \int d^5x \, \sqrt{-g_{AdS_5}} \,
\der^k\phi^* \, \der^l\phi \, F_{kp} \, F_{lq} \, g^{pq} \, ,
\label{LheuS}
\eeq
where the ten-dimensional solutions for the scalars depend on the
$S^5$ coordinates.

Note that in (\ref{LheuS}) all  fields are evaluated at the same
spacetime point, namely: we have built an effective four-point
interaction. This is referred as the ultra-local approximation. In
the supergravity picture the scattering amplitude can be
schematically written in terms of the quantum mechanical operator
language as
\begin{equation}
{\cal{A}} \sim \kappa_5^2 \sum_x \sum_{x'} \langle
T^{\phi}|x\rangle\langle x|G| x' \rangle \langle x' | T^{A} \rangle
\sim \kappa_5^2 \sum_x \langle T^{\phi}|x\rangle\langle x|G |  T^{A}
\rangle \, ,
\end{equation}
where expressions of the form $G_{\mathrm{grav}}(x,x')$ correspond
to the matrix elements $\langle x|G|x' \rangle$. Now, for the
solutions that we have described in the introduction the
ten-dimensional curved space Mandelstam variables act as second
order differential operators defined by
\begin{eqnarray}
\tilde{s}_5 &=& z^2 s + 
 \nabla_{z(s)}^{2} \, , \label{st1}\\
\tilde{t}_5 &=& z^2 t + 
\nabla_{z(t)}^{2}
\, . \label{st2}
\end{eqnarray}
In the graviton propagator $G_{\textrm{grav}}\sim \tilde{t}_5^{-1}$,
however the full the string theory pre-factor we included in the
previous paragraph depends on both $\tilde{t}_5$ and $\tilde{s}_5$.
In the DIS regime at strong coupling, the latter can be thought of
as a number instead of an operator since the second term in the
r.h.s. of equation (\ref{st1}) can be neglected with respect to the
first one. However, this is not the case for $\tilde{t}_5$.
Nevertheless, at first order there is no $\tilde{t}_5$-dependence in
the amplitude, due to the fact that we only have to consider the
imaginary part of ${\cal{G}}$. Thus, in this context ${\cal{G}}$ can
be thought of as function instead of a differential operator.
Therefore the amplitude can be considered local. More details will
be given in Section 4. Note that as mentioned in the introduction,
this approximation breaks down in the exponentially small-$x$ regime
where the last term in equation (\ref{st1}) cannot be neglected.

In order to obtain the structure functions from equation
(\ref{LheuS}) the on-shell effective action must be calculated by
inserting the AdS$_5$ solution for each field and carrying out the
integrals.

\subsection{Antisymmetric contributions}

Up to now we have considered the exchange of a spin-2 field because
the amplitude of the process scales as ${\tilde {s}}^j$
\cite{Cornalba:2006xk,Hatta:2009ra,Bartels:2009sc}. Now, in order to
investigate the leading order contribution to the antisymmetric DIS
structure functions at high energy it is necessary to consider the
exchange of gauge fields. The action (\ref{LheuS}) derived in the
last subsection gives the leading contribution to the symmetric
structure functions for the glueball. However, it gives no
information about the antisymmetric ones. In the case of QCD one
would not expect these structure functions to be present in the
electromagnetic DIS. For ${\cal{N}}=4$ SYM theory at $x \simeq 1$
these antisymmetric structure functions are sub-leading in
comparison with the symmetric ones $F_1$ and $F_2$. However, we can
see that the situation is different in the $x \ll 1$ regime, due to
the Chern-Simons term present in the supergravity action
(\ref{S5d0}). From this term the antisymmetric structure functions
arise when a gauge field is exchanged in the $t$-channel, instead of
the usual graviton exchange.
\begin{figure}[ht]
\begin{center}
\includegraphics[scale=0.8]{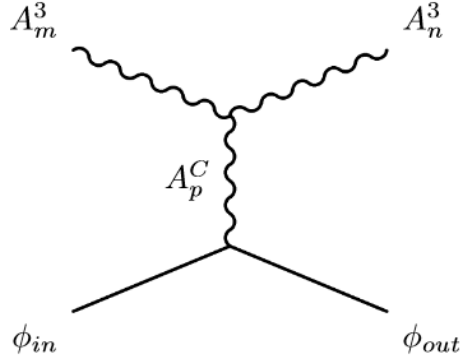}
\caption{\small  Feynman diagram corresponding to the gauge field
exchange contribution to the FCS process for small values of the
Bjorken parameter $x$.} \label{Tcs}
\end{center}
\end{figure}
Following the procedure of the Subsection 2.1, we will derive an
effective Lagrangian from which the glueball antisymmetric structure
function $F_3(x,q^2)$ can be obtained, giving a contribution of the
same order as the symmetric ones.

Since the incoming and outgoing states correspond to two $A^3_m$
gauge fields, through the Chern-Simons term they couple to another
$A^C_m$ gauge field which propagates in the AdS$_5$ space. This
state couples to two dilatons in the bulk with coupling
${\cal{Q}}^C$. In addition, there are the following eigenvalue
equations for the spherical harmonics of the dilaton\footnote{Note
that the equation below differs from the conventional eigenvalue
equation $K^a \partial_a \phi= i Q \phi$, see Appendix A. This is
due to the convention of the generators of $SU(4)$.}
\bea
K^C_a \, \der^a Y(\Omega) = - {\cal{Q}}^C \, Y(\Omega) \, ,
\label{autov}
\eea
for $K^C$ associated to the diagonal $SU(4)$ generators.

Then, the supergravity amplitude becomes
\begin{equation}
{\cal{A}} = \kappa_5^2 \int d^5x d^5x'\ {\cal{J}}^{m}_C(x) \,
G_{mn}^{CD}(x,x') \, J^{n}_{D}(x') \, , \label{acs}
\end{equation}
where ${\cal{J}}^{m}_{C}$ is the current associated with the
Chern-Simons term, while $J^n_D$ is the current associated with the
dilaton. These currents are defined as
\bea
{\cal{J}}^{m}_{C}(x) = \frac{i}{6}d_{ABC} \, \vep^{mnopq} \der_n
A_o^A \der_p A_q^B \, , \,\,\,\,\,\,\,\, \quad J^{n}_{D} (x') =  -
{\cal{Q}}_D \left( \phi\,\partial^n \phi^{*}  - \phi^{*}\partial^n
\phi \right)\ . \label{current}
\eea
Also, the gauge field propagator can be expressed as\footnote{There
is also a pure gauge component which does not contribute to this
process \cite{Bartels:2009sc}.}
\bea
G_{mn}^{CD}(x,x') = g_{mn} \, \delta^{CD} \,
G_{\textrm{gauge}}(x,x') \, .
\eea

Then, the integrand of the amplitude (\ref{acs}) becomes
\bea
{\cal{J}}^{m}_C(x) G_{mn}^{CD}(x,x')J^{n}_{D}(x') = -\frac{i}{6}
d_{ABC} {\cal{Q}}^C\ \vep^{mnopq} \der_nA_o^A \der_p A^B_q \
 \left(\phi\,\partial_{m} \phi^{*}  -
\phi^{*}\partial_{m} \phi \right) \, . \label{Lheur}
\eea
As mentioned, the incoming gauge fields correspond to photons
$A_m^3$ related to $K^3=K^3_a \, \der^a$, which is the generator of
one of the $U(1)$ subgroups of the $SU(4)_R$ group. Then, the
relevant components of the symmetric symbol are of the form
$d_{33C}$. Only $d_{338}$ and $d_{33, 15}$ contribute, which are
related to the $K^8$ and $K^{15}$ diagonal generators of $SU(4)_R$.

Now, in order to obtain the antisymmetric structure function
$F_3(x,q^2)$ in the $x\ll1$ regime we have to build an effective
Lagrangian with the tensor structure of equation (\ref{Lheur}).
Then, similarly to the symmetric case described in the previous
subsection, we must multiply by the string theory factor. The
effective Lagrangian is
\bea
S^{{\mathrm{Asym}}}_{{\mathrm{eff}}} &=& -\frac{i}{6} d_{ABC}
{\cal{Q}}^C \, {\mathrm{Im}}\left[{\cal{G}} \tilde{s}^2 \right] \,
\kappa_5^2 \,
 \times \nn \\
&&
\int d^5\Omega \, \sqrt{g_{S^5}} \, \int d^5x \, \vep^{mnopq} \,
\der_mA^{*A}_n \, \der_oA^B_p \, \left(\phi \,
\partial_{q} \phi^{*} - \phi^{*} \partial_{q} \phi \right) \, ,
\label{LheuCS}
\eea
where $A=B=3$ and $C=8,15$ for the relevant case.

Next step must be the evaluation of the effective Lagrangian on the
AdS$_5$ solutions. We present the calculation of $F_3(x,q^2)$ in
Section 4.

Although at this point ${\cal{G}}(\al',{\tilde {s}},{\tilde
{t}},{\tilde {u}})$ has been included as an {\it ad hoc} pre-factor,
it can be understood from the fact that it appears in the four-point
scattering amplitudes calculated directly from string theory. Also,
by multiplying by an extra ${\tilde {s}}^2$ we obtain an effective
Lagrangian proportional to $1/{\tilde {t}}$, which is expected when
a gauge field is exchanged in the $t$-channel. This is still an
heuristic approach. In the next section we will show explicitly how
these factors emerge from closed superstring theory scattering
amplitudes.

\section{Antisymmetric effective action from string theory}

The Lagrangian we have obtained in the previous section from the
Chern-Simons term of the five-dimensional $SU(4)$ gauged
supergravity on AdS$_5$ can be obtained from type IIB superstring
theory. Then, the {\it ad hoc} pre-factor can be straightforwardly
obtained from a first principles derivation. For that purpose we
have to calculate a four-point closed type IIB superstring theory
scattering amplitude in the ${\tilde {t}} \rightarrow 0$ limit with
insertions corresponding to two dilatons and two gauge fields $A^3$.
In the case of a graviton exchange, the gauge fields are encoded in
metric perturbations polarized in a particular way
\cite{Polchinski:2002jw}. We start from the string theory scattering
amplitude of the form ${\cal{A}}(h,h,\phi,\phi)$. String theory
scattering amplitudes include all the possible interchanged modes.
Then, a question one should ask is why the leading antisymmetric
contributions we found heuristically in the previous section cannot
be derived from ${\cal{A}}(h,h,\phi,\phi)$. The subtlety lies in the
fact that, as emphasized in \cite{Kim:1985ez}, the massless gauge
fields that appear after the $S^5$ reduction are actually linear
combinations of graviton and RR 4-form field modes. The precise
relation is given in equation (16). This means that we also have to
consider a process with ingoing RR states, {\it i.e.} a scattering
amplitude of the form
${\cal{A}}({\cal{F}}_5,{\cal{F}}_5,\phi,\phi)$.

As a consistency check, these RR modes should be associated with the
derivation of the Chern-Simons term. In the next section we will
show how it can be obtained from the amplitude
${\cal{A}}({\cal{F}}_5,{\cal{F}}_5,h)$. Then, in Subsection 3.2 we
will derive the effective Lagrangian from this term. This Lagrangian
will be used in Section 4 for the calculation of the leading
contribution to the structure function $F_3(x, q^2)$ for glueballs.

\subsection{Chern-Simons interaction from the superstring amplitude}

In this section we derive the structure of the Chern-Simons term of
five-dimensional gauged supergravity from type IIB string theory on
the AdS$_5 \times S^5$ background. Firstly, we calculate a three
closed string scattering amplitude on flat space-time, and then we
evaluate the incoming closed string states on a certain specific
{\it Ansatz}. The {\it Ansatz} corresponds to the $S^5$
compactification from the ten-dimensional type IIB supergravity
solution generating the effective $SU(4)$-gauged supergravity on
AdS$_5$ \cite{Kim:1985ez,Baguet:2015sma,Cvetic:2000nc}. In this work
we will mainly follow the first two references.

Let us focus on the {\it Ansatz} for the graviton and the RR
$4$-form field perturbations. The relevant ten-dimensional type IIB
supergravity action is given by
\bea
S_{10d}^{IIB \,\, sugra} = \frac{-1}{2\kappa^2_{10}}\int d^{10}x
\sqrt{-G}\left({\cal{R}}_{10{\mathrm{d}}}-\frac{1}{240} \, \F_5^2 \right) + \cdots \,,
  \label{s10}
\eea
together with the self-duality condition ${\cal {F}}_5 = * {\cal
{F}}_5 $, where $*$ is the Hodge dual operator in ten dimensions and
$G_{MN}$ is the ten-dimensional metric. Recall that $2 \kappa_{10}^2
= {\rm{Vol}} \left(S^5\right) 2 \kappa_5^2$ with
${\rm{Vol}}\left(S^5\right)=\pi^3$. In this notation the five-form
field strength and the self-duality condition are written as
\beq
\F_{M_1...M_5} = 5 \, \der_{[M_1} a_{M_2...M_5]} \ ,
\,\,\,\,\,\,\,\,\,\, (*\F)_{M_1...M_5}=\frac{1}{5!}\sqrt{-G} \,
\vep_{M_1...M_5N_1...N_5} \, \F^{N_1...N_5} \ .
\eeq
Type IIB supergravity action can be consistently reduced on $S^5$,
obtaining the five-dimensional $SU(4)$ gauged supergravity action
(\ref{s5d}). In \cite{Kim:1985ez} it was pointed out that the
linearized equations of motion of graviton and four-form field
excitations are decoupled from other fields, which means that it is
consistent to turn off all other fields and work only with these
perturbations. By expanding the fields in scalar, vector and tensor
spherical harmonics on $S^5$, it has been shown that only a
particular linear combination of the fundamental modes of both the
graviton and the four-form field gives rise to the massless vector
modes $A_m^A$ \cite{Kim:1985ez}. The form of the relevant
perturbations is given by
\beq
h_{ma} = A_{m}^B K_a^B  \ , \  a_{mabc} \sim A^{B}_m Z_{abc}^B
\label{pert}
\eeq
up to a numerical constant, and where the $K_a^B$ are the 15 Killing
vectors on $S^5$ (in other words, the lowest vector spherical
harmonics, thus giving the usual Kaluza-Klein {\it Ansatz} of the
metric components). $Z_{abc}^A$ is a pseudo-tensor on $S^5$ defined
from these Killing vectors, the volume form $\ep$ and the covariant
derivatives $\nabla_a$ as
\begin{equation}
Z_{abc}^A \equiv \ep_{abcde} \nabla^d K^{eA} \,
. \label{Zdef}
\end{equation}
The Levi-Civita tensor is given by
\bea
\ep_{abcde} = \sqrt{g_{S^5}} \,\,\vep_{abcde} \nn \ , \,\,\,\,\,\,\,
\ep^{abcde} = \frac{1}{\sqrt{g_{S^5}}}\,\, \vep^{abcde} \, ,
\eea
where $\vep$ is the totally antisymmetric symbol such that
$\vep_{12345}=\vep^{12345}=1$.

The starting point is the following flat-space three-point closed
superstring theory scattering amplitude\footnote{For details see
Appendix B.}
\begin{equation}
{\cal{A}} \sim \int \prod_{i=1}^{3}d^2z_i \, \langle
V_{{\mathrm{RR}}}^{(-\frac{1}{2},-\frac{1}{2})}(z_1,\bar{z}_1) \,
V_{{\mathrm{RR}}}^{(-\frac{1}{2},-\frac{1}{2})}(z_2,\bar{z}_2) \,
V_{{\mathrm{NSNS}}}^{(-1,-1)}(z_3,\bar{z}_3)\rangle \, ,
\end{equation}
where the vertex operators on the two-sphere and the corresponding
conventions can be found for example in
\cite{Bakhtiarizadeh:2013zia, Garousi:1996ad, Becker:2015eia} and
references therein. In the case we are interested in, the RR modes
correspond to self-dual five-form field strength perturbations while
the NSNS mode is the graviton. The expression has been explicitly
obtained in \cite{Becker:2015eia}
\begin{equation}
{\cal{A}}({\cal{F}}^{(1)}_5,{\cal{F}}^{(2)}_5,h) = -\frac{2i\kappa_{10}}{3} \, h^{MN} {\cal{F}}^{(1)}_{\,\,M M_1 \dots M_4}
{\cal{F}}_{N}^{(2) \,\, M_1 \dots M_4} \, ,
\end{equation}
and it corresponds to an interaction term in the type IIB
supergravity action which can be obtained by perturbing the
${\cal{F}}^2_5$ term using $G_{MN} \rightarrow G_{MN} +  h_{MN}$.

Now, the extension of this term to the curved spacetime background
can be written as
\beq
\frac{1}{3 \, \kappa_{10}^{2}} \int_{AdS_5 \times S^5} d^{10}x \,
\sqrt{-G} \, h^{MN} {\cal{F}}_{M M_1 \dots M_4}
{\cal{F}}_{N}^{\,\, M_1 \dots M_4} \, .
\eeq
By plugging the perturbations (\ref{pert}) in the above equation, it
is easy to see that the result has the following structure
\beq
\sqrt{-G} \F_{M M_1 \dots M_4}\F_{N}^{\,\,\,\,M_1 \dots M_4}h^{MN} \sim
 \left[\vep^{mnopq}\der_m A_n^A \der_o A_p^B A_q^C\right]
\left[\sqrt{g_{S^5}}\ep_{abcde} \D^a K^b_A \D^c K^d_B K^e_C \right].
\eeq
Thus, from the ten-dimensional point of view the five-dimensional
Chern-Simons term on AdS$_5$ comes with an integral over $S^5$. The
explicit computation of this integral leads to the symmetric symbol
$d_{ABC}$. For that we use equation (\ref{idep2}) given in the
Appendix A of the present work.

\subsection{The $A + \phi \rightarrow A + \phi$ scattering amplitude}

The results in the previous section indicate that in order to
calculate the effective Lagrangian (\ref{LheuCS}) there are two
relevant contributions to the $A + \phi \rightarrow A + \phi$
scattering amplitude, and particularly we need the one coming from
the massless RR-RR-NSNS-NSNS four-point closed string scattering
amplitude ${\cal{A}}({\cal{F}}_5,{\cal{F}}_5,\phi,\phi)$ obtained in
type IIB superstring theory. For small values of the Bjorken
parameter we have to focus on the ${\tilde {t}} \rightarrow 0$
limit.

The scattering amplitude is given by the worldsheet correlation
function\footnote{In this section we use the standard convention for
string theory scattering amplitudes where all external states are
ingoing. We will switch back to the $A + \phi \rightarrow A + \phi$
notation in the next section, where two of the states will be taken
to be outgoing.} of four vertex operators
\begin{equation}
{\cal{A}} \sim \int \prod_{i=1}^{4}d^2z_i \, \langle
V_{{\mathrm{RR}}}^{(-\frac{1}{2},-\frac{1}{2})}(z_1,\bar{z}_1)
V_{{\mathrm{RR}}}^{(-\frac{1}{2},-\frac{1}{2})}(z_2,\bar{z}_2)
V_{{\mathrm{NSNS}}}^{(-1,-1)}(z_3,\bar{z}_3)
V_{{\mathrm{NSNS}}}^{(0,0)}(z_4,\bar{z}_4)\rangle \, .
\end{equation}
Details of the computation can be found in
\cite{Bakhtiarizadeh:2013zia}. The final result in the case where
the RR-modes correspond to five-form field strength perturbations
and the NSNS-modes correspond to the dilaton, is given by the
product
\bea
{\cal{A}}({\cal{F}}_5^{(1)},{\cal{F}}_5^{(2)},\phi^{(3)},\phi^{(4)})
= {\cal{G}} \, {\cal{K}} \, ,
\eea
with $\tilde{s}=-2 k_1\cdot k_4$, $\tilde{t}=-2 k_1\cdot k_2$ and
$\tilde{u}=-2 k_1\cdot k_3$ the ten-dimensional Mandelstam variables
($k_1 + k_2 + k_3 + k_4 = 0$ and $\tilde{s} + \tilde{t} + \tilde{u}
= 0$), and ${\cal{K}}$ is the kinematic factor
\begin{equation}
{\cal{K}} = -80 \kappa_{10}^2 \tilde{s} \, \tilde{u} \,\phi_3 \,
\phi_4 \, {\cal{F}}^{(1)}_{\,\,M M_1 \dots M_4} {\cal{F}}_{N}^{(2)
\,\, M_1 \dots M_4} k_4^M k_4^N. \label{KFFdd}
\end{equation}
For the small-$x$ regime and within the ultra-local approximation,
we are interested in considering the small-$\tilde{t}$ limit (which
is trivial for this particular ${\cal{K}}$ except for the fact that
we can take $\tilde{u}=-\tilde{s}$) and constructing an effective
four-point interaction Lagrangian that reproduces this scattering
amplitude. The Lagrangian associated with ${\cal{A}}={\cal{G}} \,
{\cal{K}}$ in the Einstein frame turns out to be
\begin{equation}
{\cal{L}}_{{\cal{F}}_5{\cal{F}}_5\phi\phi} =
-20\kappa_{10}^2\left[
{\cal{G}}(\al',\tilde{s},\tilde{t}\rightarrow 0,\tilde{u})
\,\tilde{s}^2\right] {\cal{F}}_{M M_1 \dots M_4} \,
{\cal{F}}_{N}^{\,\,\,\, M_1 \dots M_4} \, \der^{(M}\phi
\,\,\der^{N)}\phi \,. \label{LFFdd}
\end{equation}
Finally, the full effective action written in terms of the gauge
fields and the Killing vectors associated with the expansion on
$S^5$ is obtained by writing the curved-space version of the
effective action corresponding to (\ref{LFFdd}) and inserting the
explicit form of the ${\cal{F}}_5$ perturbations (\ref{pert}). This
yields an integrand of the form
\begin{equation}
\left[ {\cal{G}}(\al',\tilde{s},\tilde{t}\rightarrow 0,\tilde{u})
\,\tilde{s}^2\right]\, \sqrt{g_{S^5}} \left(\ep^{abcde}
\D_a K_b^A \D_c K_d^B \right)  \left(\vep^{mnopq}\der_m A_n^A \der_o
A_p^B \right) \der_{(e} \phi \,  \der_{q)} \phi \, . \label{Lred}
\end{equation}
By using the relation (\ref{autov}) and the Killing vector identity (\ref{idep3}) presented in Appendix A, in the ingoing/outgoing
convention we see that both the symmetric symbol $d_{ABC}$ and the
current associated with dilaton come from
\begin{equation}
\left(\ep^{abcde} \D_a K_b^A \D_c K_d^B \right) \der_{(e} \phi \,
\der_{q)} \phi^* = \frac{4 i}{R}\, \,  d_{ABC} \, K_C^{e} \der_{(e}
\phi \,  \der_{q)} \phi^* =  \frac{2 i}{R}\,d_{ABC}\, J^C_q,
\end{equation}
where $J_{m}^C$ is the second of the currents (involving dilatons)
given in equation (\ref{current}). Also, the $d_{ABC}$ factor
combined with the second parentheses of equation (\ref{Lred})
renders the Chern-Simons current ${\cal{J}}_n^D$. This means that we
obtain the structure anticipated in equation (\ref{Lheur}). These
results are in full agreement with the effective action
(\ref{LheuCS}) we predicted in Section 2.3 using heuristic
arguments. Finally, remember that for the particular process studied
in this paper we will focus on the contribution proportional to
$d_{33C}$.

\section{Antisymmetric structure function $F_3$ at small $x$}

In this section we obtain the antisymmetric structure function
$F_3(x,q^2)$ for the glueball, following the conventions of
reference \cite{Polchinski:2002jw}\footnote{Note that in
\cite{Polchinski:2002jw} the normalization of the fields is such
that the interaction term between the dilaton and the gauge field is
\bea S_{\textrm{int}}=i Q^C\int d^{10} x
\sqrt{g}A_C^m\left(\phi^*\der_m\phi -\phi\der_m \phi^* \right). \nn
\eea}. We recover $R$ factors wherever it corresponds. As explained
in the introduction, in the $e^{-\sqrt{\lambda}} \ll x \ll
\lambda^{-1/2}$ regime the holographic method consists in evaluating
the on-shell amplitude associated with the effective supergravity
process and taking its imaginary part. Then, if we separate the
hadronic tensor into its symmetric and antisymmetric parts as
$T^{\mu\nu}=T^{\mu\nu}_{sym}+iT^{\mu\nu}_{asym}$ (and the same for
$W^{\mu\nu}$) the AdS/CFT dictionary implies
\cite{Gao:2009ze,Hatta:2009ra}
\begin{equation}
-iS^{{\mathrm{Asym}}}_{\mathrm{eff}} \equiv n_\mu n^*_\nu
\,{\mathrm{Im}} \left(T^{\mu\nu}_{{\mathrm{Asym}}}\right) =
\frac{1}{2\pi} n_\mu n^*_\nu W^{\mu\nu}_{{\mathrm{Asym}}} \, ,
\end{equation}
where the last equality follows from the optical theorem. The
calculation of $F_3$ is similar to the one corresponding to the
symmetric structure functions $F_1$ and $F_2$ presented in
\cite{Polchinski:2002jw}. The starting point is the effective action
proposed in Section 2 from heuristic arguments and derived from
first principles in Section 3. Considering two ingoing states and
two outgoing states, this on-shell action is given by
\bea
S^{{\mathrm{Asym}}}_{{\mathrm{eff}}} &=& i\frac{R}{6}\, d_{33C}
{\cal{Q}}^C \,
{\mathrm{Im}}\left[{\cal{G}}(\al',\tilde{s},\tilde{t}\rightarrow
0,\tilde{u})
\,\tilde{s}^2\right] \times \nn \\
&& \int \, d^5\Omega  \,\sqrt{g_{S^5}} \int \, d^5x \,
\vep^{mnopq} \der_m A^{3*}_n\der_o A^3_p \left(
\phi\,\partial_{q} \phi^{*} - \phi^{*}\partial_{q} \phi  \right) \,
. \label{Sfinal}
\eea
The AdS$_5$ solutions we have to
insert are given by (\ref{solphi}) and (\ref{solA}). Also, let us
recall that the relation between the ten-dimensional invariant
$\tilde{s}$ and the four-dimensional one is
\begin{equation}
\al'\tilde{s} \approx \al' s\frac{z^2}{R^2} \, ,
\end{equation}
in the regime under consideration and up to corrections from the
radial and $S^5$ components of order $\al'/R^2 \sim \lambda^{-1/2}$
which can be neglected.

As in the symmetric case, by taking the $\tilde{t}\rightarrow 0$
limit, the imaginary part of the pre-factor can be replaced by a sum
over excited string states \cite{Polchinski:2002jw}. The exact
result is
\begin{equation}
{\mathrm{Im}}_{{\mathrm{exc}}}\left[{\cal
{G}}(\al',\tilde{s},\tilde{t},\tilde{u})\
\tilde{s}^2\right]|_{\tilde{t}\rightarrow 0} = \frac{\pi
\al'}{4}\sum_{m=1}^{\infty} \delta \left(m - \frac{\al'
\tilde{s}}{4}\right) (m)^{\frac{\al' \tilde{t}}{2}}, \label{IMG}
\end{equation}
where the last factor can be ignored if $x$ is not exponentially
small, {\it i.e.} when $e^{-\sqrt{\lambda}} \ll x \ll
\lambda^{-1/2}$. This sum can be expressed in terms of $\omega = q
z$ as
\bea
\sum_{m} \delta\left(m-\frac{\al'\tilde{s}}{4}\right) =
\sum_{\omega_m} \left(\frac{2q^2R^2}{\al's \,
\omega}\right)\delta\left(\omega-\omega_m\right) \ , \,\,\,\,\,\,\,
\omega_m^2 \equiv m \frac{2R^2q^2}{ \al's} \, , \label{ImG2}
\eea
which is well approximated by an integral for $x\ll \lambda^{-1/2}$.

Plugging the solutions for the gauge fields and the dilaton current
together with equation (\ref{ImG2}) in the on-shell effective action
(\ref{Sfinal}), and working out the integration over the full
ten-dimensional spacetime, we find
\begin{equation}
n_\m n_\n^* W^{\m\n}_{asym} = \,|c'_i|^2\frac{\pi^2}{3}  \,  \,
\frac{{\cal{Q}} \,{\cal{I}}_\Delta}{\sqrt{4\pi\lambda}}
\left(\frac{\Lambda^2}{q^2}\right)^{\Delta-1} n_{\mu}n_{\n}^{*} \,
\varepsilon^{\m\n\rho\sigma} \frac{q_{\rho} P_{\sigma}}{2P\cdot q}
\frac{1}{x}\, \label{Wfinal}
\end{equation}
where the charge is ${\cal{Q}} = d_{33C}{\cal{Q}}^C$, while
${\cal{I}}_\Delta$ stands for the  $\omega$ integral over the Bessel
functions
\begin{equation}
{\cal{I}}_\Delta = \int d\omega \, \omega^{2\Delta+2} \, K_0(\omega)
\, K_1(\omega) = \frac{\sqrt{\pi}}{4}
\frac{\Gamma\left(\Delta+1\right)^2
\Gamma\left(\Delta+2\right)}{\Gamma\left(\Delta+\frac{3}{2}\right)}
\, .
\end{equation}
Now, by comparison of equation (\ref{Wfinal}) with the general
decomposition (\ref{Wdecomp}) we obtain the antisymmetric structure
function for the glueball
\begin{equation}
F_3(x,q^2) = \frac{1}{x}\, \left(\frac{\Lambda^2}{q^2}\right)^{\Delta-1}
\frac{{\cal{Q}}\,|c'_i|^2\pi^2}{3\sqrt{4\pi\lambda}}\,{\cal{I}}_\Delta
\,.
\label{F3}
\end{equation}
Let us recall that for the dilaton in the
${\mathrm{exp}}(-\sqrt{\lambda})\ll x \ll \lambda^{-1/2}$ regime,
one obtains the following symmetric structure functions
\begin{equation}
F_1(x,q^2) =
\frac{1}{x^2}\left(\frac{\Lambda^2}{q^2}\right)^{\Delta-1}
\frac{\pi^2 \rho_{\Omega}
|c_i|^2}{4\sqrt{4\pi\lambda}}I_{1,2\Delta+3}
 \,,\qquad
F_2(x,q^2) = 2x \frac{2\Delta+3}{\Delta+2}F_1(x,q^2)\,,
\end{equation}
were $\rho_{\Omega}$ is a dimensionless constant coming from the
angular integral of the symmetric effective action, defined in
equation (88) of \cite{Polchinski:2002jw}, and
\begin{equation}
I_{1,2\Delta+3} = \int d\omega \, \omega^{2\Delta+3} \,
K_1^2(\omega)\,
=\frac{(2\Delta+2)(\Delta+2)}{2\Delta+3}{\cal{I}}_\Delta \, .
\end{equation}
We find new Callan-Gross like relations that can be expressed as:
\bea
F_3(x,q^2) = \frac{{\cal{Q}}}{\rho_{\Omega}}\frac{4}{3}(\Delta+1)
F_2(x,q^2) =
\frac{{\cal{Q}}}{\rho_{\Omega}}\frac{8}{3}\frac{(2\Delta+3)(\Delta+1)}{(\Delta+2)}
x F_1(x,q^2)\, .
\eea
One subtle difference arises from the ${\cal{Q}}$ factor: for $F_3$
to be non-vanishing the hadron must be charged under the $U(1)$
groups associated to the $T_8$ and $T_{15}$ generators, while this
is not necessary for the symmetric functions.

Note that our result of equation (\ref{F3}) is in agreement with the
behavior found in \cite{Hatta:2007he} for the spin-$1/2$ case given
by a dilatino mode. In the mentioned work, the antisymmetric
structure functions are computed in the exponentially small-$x$
regime, but one can extrapolate the result by considering the
ultra-local approximation.

\subsection{Comments on the exponentially small-$x$ regime}

In the exponentially small-$x$ regime the ultra-local approximation
does not hold due to diffusion effects in the radial direction of
AdS$_5$ become important. This happens because the last factor in
equation (\ref{IMG}) cannot be neglected. Thus, one must consider
the full differential operator of equation (\ref{st2})
\cite{Polchinski:2002jw,Brower:2006ea,Brower:2007qh,Brower:2007xg,
Brower:2010wf,Koile:2014vca}.

In the symmetric case, this leads to the interchange of a Pomeron.
Let us start by reviewing this in the conformal limit. The
differential operator acts on $\partial_\mu \phi \,
\partial_\nu \phi^*$. More concretely, it is given by the spin-$2$
Laplacian, and the exponent reads
\begin{equation}
\frac{\al'\tilde{t}}{2} = \frac{1}{2}\frac{\al'}{R^2} z^2 t +
\frac{1}{2}\frac{\al'}{R^2} \Delta_2 =\frac{1}{2}
\frac{\al'}{R^2}z^2 t + \frac{1}{2\sqrt{\lambda}} \left[z^2 \der_z^2
+ z \der_z - 4\right]. \label{t2}
\end{equation}
We will set $t=0$. $\Delta_2$ is a particular case of the Hodge-de
Rham operator, defined more generally by
\beq
R^2 \Delta_j = z^2 \der_z^2 + (2j-3) z \der_z + j (j-4) \, .
\eeq
It can be evaluated in terms of an auxiliary quantum mechanical
problem where $u=-\log (z/z_{{\mathrm{ref}}})$ plays the role of
time and $H=-z^2 \der_z^2 - z \der_z + 4=-\partial_u^2 + 4$ is the
Hamiltonian. In the conformal case $z_{{\mathrm{ref}}}$ is an
arbitrary scale and there is no cut-off in the AdS spacetime. One
can then diagonalize this operator in terms of its eigenfunctions,
which are plane waves in $u$ with energies $E_\nu= \nu^2+4$. Then,
the scattering amplitude can be written in terms of a kernel which
in the $\tilde{t} \rightarrow 0$ limit takes the form
\begin{equation}
{\cal{K}}(u,u',t=0,j=2) = \left(\al' \tilde{s}\right)^{2-
\frac{2}{\sqrt{\lambda}}} \sqrt{\frac{\lambda^{1/2}}{2\pi
\tau}}e^{-\frac{\sqrt{\lambda}}{2\tau}(u-u')^2} \, ,
\end{equation}
where $\tau = \log (\alpha' \tilde{s})$. Note that the $s^2$ factor
was already present in the ultra-local approximation of the
scattering amplitude. It reflects the appropriate scaling with the
center-of-mass energy for a graviton exchange. The $(u-u')^2$
dependence in the exponential is known as the diffusion factor and
the inverse of its coefficient gives the associated characteristic
diffusion time. The final DIS amplitude is obtained by evaluating
the rest of the gauge-field part of the effective Lagrangian at $u$
and the $\phi$-field part at $u'$, and integrating. For example, the
part of the on-shell effective action that contributes to the
$F_1(x,q^2)$ structure function reads\footnote{This expression is
valid after the angular integration on $S^5$. Also the scalar
solution $\phi$ does not include the scalar spherical harmonic.
Finally, notice that in these expressions we have absorbed a factor
of $4\pi$ in the definition of the 't Hooft coupling $\lambda$. }
\begin{eqnarray}
n_\mu n_\nu^* T^{\mu\nu}_{\mathrm{Sym}}|_{F_1}&= &\frac{\pi \alpha'
\rho_{\Omega} R^2}{2} \int dz  \sqrt{-G} F^{mn}F^{p*}_{\,\,n}|_{F_1}
\left(\frac{\alpha' \tilde{s}}{4}\right)^{\alpha' \tilde{t}/2|_{t=0}} \partial_m \phi^* \partial_p \phi \nn \\
&=& \frac{\pi \rho_{\Omega}}{8} \lambda^{1/2} \int \frac{dz}{z} \left[q z K_1(q z)\right]^2 \left(\al' \tilde{s}\right)^{2+\alpha' \tilde{t}/2|_{t=0}}\left[\frac{R^4}{z}|\phi(z)|\right]^2\, \nn \\
&=& \frac{\pi \rho_{\Omega}}{8} \lambda^{1/2} \int \frac{dz}{z} \frac{dz'}{z'} \left[q z K_1(q z)\right]^2 \left(\al' \tilde{s}\right)^{2+\frac{1}{2\sqrt{\lambda}}\Delta_2}\delta (u(z) - u'(z'))\left[\frac{R^4}{z'}|\phi(z')|\right]^2\, \nn \\
&\equiv& \frac{\pi \rho_{\Omega}}{8} \lambda^{1/2} \int du \, du'
P_A^{(1)}(u) {\cal{K}}(u,u',t=0,j=2) P_\phi(u'), \label{F1F2pomeron}
\end{eqnarray}
where $P_A^{(1)}(u(z))=  q^2 z^2 K_1^2(q z)$ and $P_\phi(u(z))= R^8
z^{-2} |\phi(z)|^2 \approx (z\Lambda)^{2\Delta-2}$ are scalar
factors that only depend on the corresponding incoming solutions,
and all contractions are made with the curved metric. In the last
step we have written everything in terms of $u$ and inserted an
identity of the form $\int du' \delta(u-u') = \int du' \int
\frac{d\nu}{2\pi} e^{i \nu (u-u')} $, which naturally leads to the
appearance of the spin-2 kernel. Now, due to the optical theorem
$F_1(x,q^2)$ is obtained simply by multiplying by a $2\pi$ factor. A
similar expression can be found for $F_2 (x,q^2)$ by replacing
$P_A^{(1)} \rightarrow P_A^{(2)}(u(z))= q^2 z^2 (K_0^2(q z)+K_1^2(q
z))$ and inserting an extra factor of $2x$. Of course, in these
final formulas the $x$-dependence is hidden in the $s$ and $\tau$
factors since in this regime the four-dimensional Mandelstam
invariant is $s \approx q^2/x$. Also, the result for the parametric
region $\exp{(-\lambda^{1/2})} \ll x \ll {\lambda}^{-1/2}$ is
formally recovered in the large $\lambda$ limit.

When the cut-off at $z_0$ is introduced in the AdS$_5$ spacetime to
induce confinement general steps of the above calculation remain
valid, but one has to impose boundary conditions on $z_0$,
consistent with energy-momentum conservation. Taking the reference
value as $z_{{\mathrm{ref}}}=z_0$, the boundary condition on the
Pomeron modes $h_{++}$ \footnote{We are using light-cone
coordinates. Also, we consider only modes which are relevant in the
high energy limit.} and the resulting eigenfunctions are
\beq
\der_z(z^2 h_{++})|_{z_0} = 0 \ \Rightarrow \ h_\nu(u) =
\frac{1}{\sqrt{2}} \left[ e^{i \nu u} + \left(\frac{\nu - 2i}{\nu +
2i}\right) e^{-i \nu u}  \right].
\eeq
Therefore, in the $t \rightarrow 0$ limit the conformal kernel must
be replaced by
\begin{equation}
{\cal{K}}_{\Lambda}(u,u',t=0,j=2) = \left(\al' \tilde{s}\right)^{2-
\frac{2}{\sqrt{\lambda}}} \sqrt{\frac{\lambda^{1/2}}{2\pi
\tau}}\left[e^{-\frac{\sqrt{\lambda}}{2\tau}(u-u')^2} +
F(u,u',\tilde{\tau})
e^{-\frac{\sqrt{\lambda}}{2\tau}(u+u')^2}\right]
%
\end{equation}
where $\tilde{\tau} = (4\lambda)^{-1/2} \tau$ and we have defined
the function
\begin{equation}
F(u,u',\tilde{\tau}) = 1-4\sqrt{\pi \tilde{\tau}}e^{\eta^2}
{\mathrm{erfc}}(\eta) \ ,
\end{equation}
with
\begin{equation}
\eta = \frac{u+u'+4\tilde{\tau}}{\sqrt{4\tilde{\tau}}} \ , \
{\mathrm{erfc}}(\eta) = 1-{\mathrm{erf}}(\eta) =
\frac{2}{\sqrt{\pi}} \int_\eta^\infty dx \, e^{-x^2}.
\end{equation}
Note that $-1<F(u,u',\tilde{\tau})<1$. These results are important
to understand the holographic DIS process at high energies. In fact,
the structure of the amplitude at strong coupling written in terms
of the Pomeron kernel has a striking formal resemblance with the one
obtained at weak coupling. Also, the comparison with the available
data for DIS at small $x$ leads to some very interesting results
\cite{Brower:2010wf}.

The process we have been analyzing is such that the leading
contribution to the $F_3$ antisymmetric structure function comes
from the exchange of a Reggeized gauge field. This was also pointed
out in the DIS from dilatinos in reference
\cite{Hatta:2009ra}\footnote{Since the analysis of
\cite{Hatta:2009ra} is similar to what we need for the dilaton case,
we only outline the important steps.}. As we have seen in Section 2,
this vector mode interacts with the currents instead of the
energy-momentum tensors, implying that we have to consider the
spin-one differential operator. Thus, in this case we have
\begin{equation}
\frac{\al'\tilde{t}}{2}  = \frac{1}{2}\frac{\al'}{R^2} z^2 t +
\frac{1}{2}\frac{\al'}{R^2} (\Delta_1+3) =
\frac{1}{2}\frac{\al'}{R^2} z^2 t + \frac{1}{2\sqrt{\lambda}}
\left[z^2 \der_z^2 - z \der_z\right]. \label{t1}
\end{equation}
By introducing $\rho = 2u = -2\ln(z/z_{{\mathrm{ref}}})$ we can
rewrite $\Delta_1+3=4 (\der_\rho^2 + \der_\rho)$. After
diagonalization of the relevant operator, we obtain a conformal
kernel of the form
\begin{equation}
{\cal{K}}(\rho,\rho',t=0,j=1) = \left(\al'
\tilde{s}\right)^{1-\frac{1}{2\sqrt{\lambda}}} e^{- \frac{1}{2}
(\rho + \rho')} \sqrt{\frac{\lambda^{1/2}}{2\pi \tau}}
e^{-\frac{\sqrt{\lambda}}{8\tau}(\rho-\rho')^2}.
\end{equation}
The Regge slope is now $1-1/(2\sqrt{\lambda})$ since both the
scaling with the center-of-mass and its curvature correction change.
Note that this implies that $F_3(x,q^2)$ grows more rapidly as $x
\rightarrow 0$. Also, diffusion in $\rho(z)$ is still present. Now,
let us consider the effect of introducing a cut-off in the AdS$_5$
spacetime. The boundary condition on the Reggeized gauge field modes
$A_+$ is
\beq
\der_z \left(z A_+\right)|_{z_0}=0 \, .
\eeq
However, the eigenfunctions are modified in such a way that this
condition is actually analogous to the one above, leading to an
identical modification of the kernel. Therefore, we obtain
\begin{equation}
{\cal{K}}_{\Lambda}(\rho,\rho',t=0,j=1) =
\left(\al'\tilde{s}\right)^{1-\frac{1}{2\sqrt{\lambda}}}
e^{-\frac{1}{2}(\rho+\rho')} \sqrt{\frac{\lambda^{1/2}}{2\pi \tau}}
\left[e^{-\frac{\sqrt{\lambda}}{8\tau}(\rho-\rho')^2}
+F(\rho/2,\rho'/2,\tilde{\tau})
e^{-\frac{\sqrt{\lambda}}{8\tau}(\rho+\rho')^2}\right]
%
%
\end{equation}
The final form of the structure function in this regime is given by
complicated integrals in $\rho$ and $\rho'$. The formal expression
obtained for $F_3$ can be split in the conformal
$F_3^{{\mathrm{conformal}}}$, {\it i.e.} from the complete AdS$_5$
spacetime with no IR cut-off, plus the contribution from the
deformation induced by the IR cut-off:
\bea
F_3 = F_3^{{\mathrm{conformal}}} + F_3^{{\mathrm{deformation}}} \, .
\eea
The explicit result is obtained by following the same steps that led
us to equation (\ref{F1F2pomeron}), together with the insertion of
the hard-wall spin-1 kernel instead of the conformal one. Thus,
$F_3(x,q^2)$ is given by
\bea
F_3(x,q^2) &=& \frac{{\cal{Q}}\pi^2}{12} \int d\rho\,d\rho' \, {\cal{P}}_A(\rho,q) \, \times \label{F3pomeron} \\
&& \left\{ \left(\al' \tilde{s}\right)^{1-\frac{1}{2\sqrt{\lambda}}}
\,
\sqrt{\frac{\lambda^{1/2}}{2\pi\tau}} \left[
e^{-\frac{\sqrt{\lambda}}{8\tau}(\rho-\rho')^2} +
F(\rho/2,\rho'/2,\tilde{\tau})
e^{-\frac{\sqrt{\lambda}}{8\tau}(\rho+\rho')^2} \right]\right\}
P_\phi (\rho',\Lambda)    \, , \nn
\eea
where
\begin{equation}
{\cal{P}}_A(z,q) = q^3  z^3 K_0(q z) K_1(q z) \ , \,\,\,\,\,\,\,
P_\phi(z',\Lambda)= R^8 z'^{-2} |\phi(z')|^2 .
\end{equation}
The information about the $x$-dependence is contained in the
$\tilde{s}$ and $\tau$ factors. Note that the contribution from the
IR cut-off is model dependent, in the sense that it is sensible to
how the AdS$_5$ space is deformed near $z_0$. However, the conformal
contribution is model independent.

It is difficult to obtain an analytical expression for the above
integral (\ref{F3pomeron}). However, some approximations obtained
from simplification of the scalar and gauge field external solutions
lead to interesting results \cite{Brower:2010wf}. In that reference
the wave-functions products were approximated by Dirac delta
functions determined by the relevant scales
\begin{equation}
{\cal{P}}_A (z,q) \approx \frac{1}{q}\delta \left(z -
\frac{1}{q}\right) \ , \,\,\,\,\,\, P_{\phi}(z') \approx
\frac{1}{M} \delta\left(z' - \frac{1}{M}\right) \, ,
\end{equation}
where $M$ stands for some relevant mass scale, for example the
proton mass. It should be directly related to the IR cut-off scale
identified as $\Lambda \sim \Lambda_{{\mathrm{QCD}}}$. With this
approximation, the final expression for the conformal contribution
takes the form
\begin{equation}
  F_3(x,q^2) \approx \frac{{\cal{Q}}\pi^2}{3} (\alpha'\tilde{s})^{1-\frac{1}{2\sqrt{\lambda}}}
   \sqrt{\frac{\lambda^{1/2}}{2\pi\tau}} e^{-\frac{\sqrt{\lambda}}{2\tau}\log^2(q/M)} \approx
   \frac{1}{x}\frac{{\cal{Q}}\pi^2}{3\sqrt{\lambda}} \left(\frac{q}{M}\right)
   (\alpha'\tilde{s})^{-\frac{1}{2\sqrt{\lambda}}}
   \sqrt{\frac{\lambda^{1/2}}{2\pi\tau}} e^{-\frac{\sqrt{\lambda}}{2\tau}\log^2(q/M)}, \label{F3deltas}
\end{equation}
while the hard-wall contribution can be obtained using a similar
simplification. Note that in the context of the kernel notation
$\tilde{s}$ (and $\tau$) should be thought of as symmetrized in $z$
and $z'$, and in (\ref{F3deltas}) this implies
$\alpha'\tilde{s}=\frac{s}{\sqrt{\lambda}} \frac{1}{q M} =
\frac{1}{\sqrt{\lambda}}\frac{1}{x} \frac{q}{M}$
\cite{Brower:2006ea,Brower:2010wf}.

\section{Discussion}

In this work we describe how the antisymmetric structure function
$F_3(x,q^2)$ is obtained in the dual holographic description of DIS
of charged leptons from glueballs at small-$x$ in ${\cal{N}}=4$ SYM
theory deformed by the introduction of the IR scale $\Lambda$. The
reason for the non-vanishing $F_3(x,q^2)$ comes from the chiral
anomaly of ${\cal{N}}=4$ SYM theory, which does not depend on the IR
deformation. This anomaly can be seen from the three-point
correlation function of current operators, and is proportional to
the symmetric symbol $d_{ABC}$. From the string theory point of view
this comes from the $S^5$ dimensional reduction of type IIB
supergravity, which leads to the $SU(4)$ gauged supergravity on
AdS$_5$. Its action contains Chern-Simons term proportional to the
symbol $d_{ABC}$. Thus, there is a deep connection between the
chiral anomaly of the ${\cal{N}}=4$ SYM theory and the emergence of
$F_3(x,q^2)$. On the other hand, the fact that the chiral anomaly is
related to the Chern-Simons term in the bulk is reflected in the
fact that $F_3(x,q^2)$ has the power dependence in the Bjorken
variable which comes from the propagation of a gauge field in the
$t$-channel Feynman-Witten diagram of $SU(4)$ gauged supergravity in
the bulk.

In QCD $F_3$ is zero for the electromagnetic DIS, {\it i.e.} a
charged lepton scattered from a hadron with exchange of a virtual
photon, due to the fact that this particular structure function does
not preserve the parity symmetry. Of course, this would not be the
case when considering an interaction mediated by a $W^{\pm}$ or
$Z^0$ gauge boson such as in neutrino DIS. However, though QCD and
IR-deformed ${\cal{N}}=4$ SYM theories may share some generic
features in the large-$N$, these gauge theories are essentially
different. In particular, ${\cal{N}}=4$ SYM theory is chiral. The
R-symmetry current associated with the global $U(1)_R \subset
SU(4)_R$ can be gauged in order to describe the {\it electric}
current, therefore allowing for the construction of a bulk dual
photon which mediates the DIS process.  The $F_3$ structure function
was not analyzed in the original calculation developed in
\cite{Polchinski:2002jw}, but it was taken into account to some
extent in related papers such as \cite{Gao:2009ze,Hatta:2009ra} from
a heuristic viewpoint for the case of spin-1/2 hadrons.

In the supergravity regime, {\it i.e.} when $\lambda^{-1/2} \ll x
<1$, the amplitude is dominated by the $s$-channel diagram and the
corresponding contributions to $F_3$ are sub-leading in comparison
with the symmetric structure functions $F_1$ and $F_2$ for
glueballs. However, for polarized spin-$1/2$ hadrons $F_3=F_2=2 F_1$
due to the form of the associated AdS$_5$ solutions
\cite{Gao:2009ze}. The scattering process in this context is the
same as for the parity-preserving structure functions $F_1$ and
$F_2$.

In the small-$x$ regime the situation changes drastically because
excited strings must be included as intermediate states. The
dominant diagrams are given by $t$-channel Reggeized particle
exchange. In the original description these modes belong to the
tower of states associated with the graviton. This leads to the
$x$-dependence for $F_1$ and $F_2$ of the form\footnote{We omit
further corrections from the Pomeron kernel.}
$x^{-2+2/\sqrt{\lambda}}$ and $x^{-1+2/\sqrt{\lambda}}$,
respectively. However, this process only gives contributions to the
structure functions which characterize the symmetric part of the
hadronic tensor $W^{\m\n}$. This is not the right place to look for
$F_3$. After the graviton exchange, the next-to-leading order
contribution in terms of center-of-mass energy scaling is given by
gauge field exchange. As we have shown, it is in this context that
the leading antisymmetric contributions appear.

As originally suggested in \cite{Hatta:2009ra}, the presence of the
cubic Chern-Simons interaction in the five-dimensional gauged
supergravity theory is crucial, as it leads to the possibility of a
gauge field exchange with the necessary four-dimensional index
structure. We have described the corresponding scattering amplitude
from two different perspectives. On the one hand, after describing
the technique in the well-known symmetric case, we have constructed
an effective local four-point interaction Lagrangian by considering
symmetry properties, starting from the five-dimensional $SU(4)$
gauged supergravity Lagrangian \cite{Gunaydin:1985cu}. In addition,
confirming our heuristic results, we have arrived to the same
effective action directly from the analysis of a four-point type IIB
superstring theory scattering amplitude. The difference with the
symmetric case comes from the fact that one needs to consider RR
vertex operators in order to include the $t$-channel Chern-Simons
contribution. This is due to the role that the ten-dimensional
self-dual five-form field strength ${\cal{F}}_5$ plays in the
construction of the gauge fields (described at the linear level in
\cite{Kim:1985ez}) when reducing the theory on $S^5$. More
specifically, in the symmetric case the relevant modes are given by
two dilaton and two graviton perturbations (with specific
polarizations), whereas in the antisymmetric case we find that the
relevant scattering amplitude is of the form
\beq
{\cal{A}}({\cal{F}}_5,{\cal{F}}_5,\phi,\phi) \, ,
\eeq
as suggested by the analysis of Section 3.1.

Focusing on the dependence in the Bjorken parameter, the precise
calculation of the amplitude leads to
\begin{equation}
F_3(x) \sim \left(\frac{1}{x}\right)^{1-\frac{1}{2\sqrt{\lambda}}}
\, . \label{F3conclusion}
\end{equation}
This means that DIS of a charged lepton from a scalar has a non-zero
$F_3$ even when it was subleading for larger values of $x$.  The
result we show in equation (\ref{F3conclusion}) leads to two
interesting conclusions. Firstly, in the small-$x$ region $F_3$ does
not vanish even for scalar hadrons. Furthermore, the first term in
the exponent implies that $F_3$ is not sub-leading since it grows as
$F_2$ does. Secondly, the second term of the exponent shows that an
${\cal{O}}(\lambda^{-1/2})$ shift appears in the exponentially
small-$x$ region as in the symmetric case. However, the differential
operator in the $t$-channel Laplacian is now associated to spin-one
fields as opposed to the spin-$2$ operator considered in
\cite{Polchinski:2002jw,Brower:2006ea}. Thus, it leads to a
different shift. The particular value is of the same sign, but it is
smaller, which means that $F_3$ actually grows faster than $F_2$ for
extremely small values of the Bjorken parameter.

In the symmetric structure functions, at some point, the fast rising
of the single-Pomeron exchange results when $x\rightarrow 0$ will
fulfil the Froissart bound. In order to restore unitarity, it is
necessary to consider the contribution from loop diagrams, {\it
i.e.} sub-leading contributions in the $1/N^2$ expansion. In the
high energy limit these contributions are dominated by multi-Pomeron
exchange. As it is known, the formalism used above can be readily
generalized to include these diagrams by using the eikonal notation.
The eikonal formula resumes the full class of ladder diagrams, where
the exchanged particles lead to the inclusion of Pomeron
propagators, build from the Pomeron kernel \footnote{In this
context, one needs both the imaginary and the real part of the
kernel.}. From these one can construct the eikonal phase $\chi$. The
saturation regime is reached when $\chi \sim 1$ \cite{Brower:2010wf,
Hatta:2007he, Cornalba:2008sp}. We think that similar features would
take place for the antisymmetric contributions studied in this
paper. However, one should be cautious in performing the eikonal
approximation for the $j=1$ exchange since there are some subtleties
that should be taken into account \cite{Cornalba:2006xk}.

~

~

\centerline{\large{\bf Acknowledgments}}

~

We thank Carlos N\'u\~nez for a critical reading on this manuscript
and comments. This work has been supported by the National
Scientific Research Council of Argentina (CONICET) and the National
Agency for the Promotion of Science and Technology of Argentina
(ANPCyT-FONCyT) Grant PICT-2015-1525.

\newpage

\appendix

\section{Conventions for the Killing vectors on $S^5$}

In this appendix we describe the explicit relation between the
$SU(4)$ Gell-Mann matrices and the $S^5$ Killing vectors.

The Lie algebra of $SU(4)$ describes the full set of $4 \times 4$
traceless hermitian matrices. The canonical basis given is by
$\{T_{A}; A = 1, \dots, 15 \} $, where for example
\begin{equation}
    T_3= \frac{1}{2}
  \left( {\begin{array}{cccc}
   1 & 0 & 0 & 0 \\
   0 & -1 & 0 & 0 \\
   0 & 0 & 0 & 0 \\
   0 & 0 & 0 & 0 \\
  \end{array} } \right), \
   T_8= \frac{1}{2\sqrt{3}}
  \left( {\begin{array}{cccc}
   1 & 0 & 0 & 0 \\
   0 & 1 & 0 & 0 \\
   0 & 0 & -2 & 0 \\
   0 & 0 & 0 & 0 \\
  \end{array} } \right), \
   T_{15}= \frac{1}{2\sqrt{6}}
  \left( {\begin{array}{cccc}
   1 & 0 & 0 & 0 \\
   0 & 1 & 0 & 0 \\
   0 & 0 & 1 & 0 \\
   0 & 0 & 0 & -3 \\
  \end{array} } \right) \, ,
\end{equation}
are the only diagonal elements. The $T$ matrices satisfy the
orthonormality condition and the commutation relations
\begin{equation}
    {\mathrm{Tr}}(T_A T_B) = \frac{1}{2} \delta_{AB} \ , \,\,\,\,\,\,\, [T_A,T_B]=i f_{ABC} \, T_C,
\end{equation}
where $f_{ABC}$ are the completely antisymmetric structure
constants. For $SU(N \geq 3)$ it is also useful to consider a
completely symmetric symbol $d_{ABC}$ which appears in the
anti-commutations. In terms of traces of the generators, these
objects are given by
\begin{equation}
     f_{ABC}=-2i \, {\mathrm{Tr}} \left(T_A [T_B,T_C]\right) \ , \
     d_{ABC}=2 \, {\mathrm{Tr}} (T_A \{T_B,T_C\}) \, .
\end{equation}
$SU(4)_R$ is the R-symmetry group of ${\cal{N}}=4$ SYM theory, and
the $d_{ABC}$ symbol appears for example in the anomaly of the
three-point function of the R-currents. The $d_{ABC}$ symbol appears
in front of the Chern-Simons interaction in the action of the dual
gravitational theory
\cite{Witten:1998qj,Freedman:1998tz,Bilal:1999ph}. In gauge/gravity
duality applications, the electromagnetic current in general is
modeled by gauging a $U(1)_R \subset SU(4)_R$, whose generator is
generally associated with $T_3$ \cite{Hatta:2009ra}. Thus, in the
{\it electromagnetic} DIS of ${\cal{N}}=4$ SYM theory, for the
antisymmetric structure functions we are only interested in the
$d_{33C}$ components. The only non-vanishing components are
$d_{338}=1/\sqrt{3}$ and $d_{33,15}=1/\sqrt{6}$.

On the other hand, in terms of the gauge/gravity duality the
R-symmetry is realized as the isometry group of the five-sphere,
$SO(6)$, which is isomorphic to $SU(4)$. In this context, one has a
different basis given by the 15 Killing vectors $K_{[ij]}$. Now, let
us consider the canonical embedding of $S^5$ into the Euclidean
space R$^6$. The Killing vectors are the rotation generators
\begin{equation}
    K_{[ij]} = x^i\der_j-x^j\der_i \ , \,\,\,\,\,\,\, i, j=1, \dots, 6 \, ,
\end{equation}
where $x^i$ are Cartesian coordinates on R$^6$. For example, the
precise mapping for the diagonal $T$ generators is
\begin{eqnarray}
&& T_3 \leftrightarrow K_3 = 2i \left(K_{[14]}+K_{[26]}\right) \ ,
\,\,\,\,\,\,\, T_8 \leftrightarrow K_8 =
\frac{i}{2\sqrt{3}}\left(K_{[14]}-K_{[26]}+2K_{[35]}\right) \ ,
\nonumber \\
&& T_{15} \leftrightarrow
K_{15}=\frac{i}{\sqrt{6}}\left(-K_{[14]}+K_{[26]}+K_{[35]}\right).
\label{Tdiag}
\end{eqnarray}
The resulting Killing vectors are normalized as\footnote{The
normalization has a minus sign due to the imaginary unit included in
equation (\ref{Tdiag}).}
\begin{equation}
    \int_{S^5} d^5\Omega \, \sqrt{g_{S^5}} \, K_A^a \, K_B^b \, g_{ab}(S^5)
    = -\frac{\pi^3R^7}{6} \, \delta_{AB} \, .
\end{equation}
Defining a new symmetric symbol as
$\tilde{d}_{[ii'][jj'][kk']} \equiv
\vep_{ii'jj'kk'}$ (which of course only takes non-zero values $\pm
1$), one has the identity
\begin{equation}
\vep_{ii'jj'kk'} = \frac{3}{4\pi^3R^6} \, \int_{S^5} d^5\Omega \,
\vep^{abcde} \, K_a^{[ii']} \, \der_b K_c^{[jj']} \,  \der_d
K_e^{[kk']} \, , \label{idep}
\end{equation}
which leads to the expression for the Chern-Simons interaction given
in \cite{Baguet:2015sma}. In the $\{ K_A \}$ basis, equation
(\ref{idep}) becomes an integral expression for $d_{ABC}$ in terms
of the five-dimensional Levi-Civita symbol and the Killing vectors
(together with their derivatives) given by
\begin{equation}
d_{ABC} = \frac{3 i}{2\pi^3 R^6} \, \int_{S^5} d^5\Omega \,
\vep^{abcde} \, K_a^A \, \der_b K_c^B \, \der_d K_e^C \, .
\label{idep2}
\end{equation}
This allows one to rewrite the Chern-Simons term in the action in
the more familiar notation.

We can also write the additional identity
\begin{equation}
\vep^{abcde} \, \der_a K_b^A \, \der_c K_d^B = \frac{4i}{R} \, d_{ABC} \, K_C^e \,,
\label{idep3}
\end{equation}
which is usefull in the computation of the effective action in
Subsection 3.2.

Note that in this language the relevant combination is $d_{33C} \,
K_C = (i/2) \, K_{[35]}$, which means that our final result for the
structure function $F_3$ is proportional to the eigenvalue
${\cal{Q}} \equiv d_{33C} \, Q_C = (1/2) \, Q_{[35]}$ of the
spherical harmonic that defines the dilaton solution with respect
rotations on the internal $(3, 5)$-plane. This means that $F_3$ is
non-vanishing for hadrons charged with respect to $K_{[35]}$.

~

\section{Gamma matrix algebra in the three-point closed string scattering amplitude}

The starting point is the flat-space three-point type IIB
superstring theory scattering amplitude ${\cal
{A}}^{(3)}\textrm{(RR,RR,NSNS)}$, involving two RR-vertex operators
and one NS-vertex operator. This is given in \cite{Becker:2015eia}
as follows
\bea
{\cal{A}}^{(3)}_{\rm{closed}}=-\frac{i}{2} \kappa_{10} \, h_{MN}^1
{\rm{Tr}}\left(\zeta^2 {\cal{C}}\Gamma^M \zeta^{3T}{\cal{C}}\Gamma^N
\right) \, , \label{A3clos}
\eea
where $\zeta$ is given by
\begin{equation}
\zeta_{AB}^{IIB}={\cal
{F}}^{(5)}_{M_1 \cdots M_5}\left({ \cal{C}}\Gamma^{M_1
\cdots M_5}\right)_{AB} \, . \label{PolF}
\end{equation}
Now, we have to calculate the trace of twelve gamma matrices. We
follow the notation of reference \cite{Garousi:1996ad}, appendix B.
The conjugation matrix ${\cal{C}}$ raises and lowers the indices of
the gamma matrices. The corresponding indices in the definition of
the gamma matrices are $\left(\Gamma^M\right)_A^{\phantom{P}B}$,
being $A$ and $B$ Dirac indices. We have the following properties
\bea
    {\cal{C}}_{AB}^{-1}{\cal{C}}^{BC} = \delta_A^C &,&     \qquad {\cal{C}}^{AB}=-{\cal{C}}^{BA} \ , \nn \\
    \left(\Gamma^M\right)_{AB} = {\cal{C}}^{-1}_{BC} \left(\Gamma^M\right)^{\phantom{P}C}_A  &,& \qquad
    \left(\Gamma^M\right)^{AB} = {\cal{C}}^{AC} \left(\Gamma^M\right)^{\phantom{P}B}_C \ , \nn \\
    {\cal{C}}^{AC}\left(\Gamma_M\right)^{\phantom{P}D}_C {\cal{C}}^{-1}_{DB} =
    - \left(\Gamma_M^T\right)_{\phantom{P}B}^A  &,& \qquad
     \left(\Gamma_M\right)^{\phantom{P}B}_A = - {\cal{C}}^{-1}_{AC} \left(\Gamma_M^T\right)_{\phantom{P}D}^C{\cal{C}}^{DB} \ , \nn
\eea
also we use the fact that for any two matrices which are themselves
products of gamma matrices
\begin{equation}
    R_{AB}S^{BC} = -R_A^{\phantom{P}B}S_B^{\phantom{P}C}\nn \, .
\end{equation}
Other useful properties are listed below
\bea
    {\cal{C}}_{AB}^{-1} = - {\cal{C}}_{BA}^{-1}\ , \qquad \left(\left(\Gamma^{M_1\cdots M_5}
    \right)^T\right)^A_{\phantom{P}B} = -{\cal{C}}^{AC}\left(\Gamma^{M_5\cdots M_1}
    \right)_C^{\phantom{P}D} {\cal{C}}^{-1}_{DB} \, .
\eea

Then, the scattering amplitude becomes
\bea
    {\cal{A}}^{(3)}_{\rm{closed}} &=&
%
-\frac{i}{2}\kappa_{10} h_{MN}^1{\cal{F}}^{2}_{M_1 \cdots M_5}
    {\cal{F}}^{3}_{N_1 \cdots N_5}{\rm{Tr}}\left(\Gamma^{M_1}\cdots\Gamma^{M_5}\Gamma^{M}
    \Gamma^{N_1}\cdots\Gamma^{N_5}\Gamma^{N}\right) \, .
\eea

Next we can calculate the last trace using similar arguments as in
appendix A of reference \cite{Bakhtiarizadeh:2013zia}. Therefore,
one obtains the following contractions:
\beq
    {\cal{A}}^{(3)}_{\rm{closed}}=-\frac{i}{15} \kappa_{10} \ h^1_{MN}\left[5 \left({\cal{F}}^M_2 \cdot {\cal{F}}^N_3+{\cal{F}}^M_3 \cdot {\cal{F}}^N_2 \right)
    - g^{MN}{\cal{F}}_1 \cdot {\cal{F}}_2\right] \, ,
\eeq
where omitted indices are contracted. The last term vanishes since
$h$ is traceless.


\end{document}